\documentclass[a4paper, preprint, superscriptaddress, nofootinbib]{revtex4-1}    
\usepackage{graphicx}
\usepackage{subfigure}
\usepackage{siunitx}
\usepackage{amsmath,amssymb,latexsym,mathrsfs}
\usepackage{hyperref}   
\usepackage{paralist}   

\begin{document}

\title{Feasibility of Determining Diffuse Ultra-High Energy Cosmic Neutrino Flavor Ratio through ARA Neutrino Observatory}


 \author{Shi-Hao \surname{Wang}}
    \email{wsh4180@gmail.com}
    \affiliation{Graduate Institute of Astrophysics, National Taiwan University, Taipei 10617, Taiwan, R.O.C.}
    \affiliation{Leung Center for Cosmology and Particle Astrophysics, National Taiwan University, Taipei 10617, Taiwan, R.O.C.}
 \author{Pisin Chen}
    \email{pisinchen@phys.ntu.edu.tw}
    \affiliation{Graduate Institute of Astrophysics, National Taiwan University, Taipei 10617, Taiwan, R.O.C.}
    \affiliation{Leung Center for Cosmology and Particle Astrophysics, National Taiwan University, Taipei 10617, Taiwan, R.O.C.}
    \affiliation{Department of Physics, National Taiwan University, Taipei 10617, Taiwan, R.O.C.}
    \affiliation{Kavli Institute for Particle Astrophysics and Cosmology, SLAC National Accelerator Laboratory, Menlo Park, CA 94025, U.S.A.}
 \author{Jiwoo Nam}
    \email{jwnam@phys.ntu.edu.tw}
    \affiliation{Graduate Institute of Astrophysics, National Taiwan University, Taipei 10617, Taiwan, R.O.C.}
    \affiliation{Leung Center for Cosmology and Particle Astrophysics, National Taiwan University, Taipei 10617, Taiwan, R.O.C.}
 \author{Melin Huang}
    \email{phmelin@snolab.ca}
    \affiliation{Leung Center for Cosmology and Particle Astrophysics, National Taiwan University, Taipei 10617, Taiwan, R.O.C.}

\begin{abstract}
    The flavor composition of ultra-high energy cosmic neutrinos (UHECN) carries precious information about the physical properties of their sources,  the nature of neutrino oscillations and possible exotic physics involved during the propagation.
    Since UHECN with different incoming directions would propagate through different amounts of matter in Earth and since different flavors of charged leptons produced in the neutrino-nucleon charged-current (CC) interaction would have different energy-loss behaviors in the medium, measurement of the angular distribution of incoming events by a neutrino observatory can in principle be employed to help determine the UHECN flavor ratio.
    %
    %
    In this paper we report on our investigation of the feasibility of such an attempt. Simulations were performed, where the detector configuration was based on the proposed Askaryan Radio Array (ARA) Observatory at the South Pole, to investigate the expected event-direction distribution for each flavor.
    Assuming $\nu_{\mu}$-$\nu_{\tau}$ symmetry and invoking the standard oscillation and the neutrino decay scenarios, the probability distribution functions (PDF) of the event directions are utilized to extract the flavor ratio of cosmogenic neutrinos on Earth.
    %
    %
The simulation results are summarized in terms of the probability of flavor ratio extraction and resolution as functions of the number of observed events and the angular resolution of neutrino directions. We show that it is feasible to constrain the UHECN flavor ratio using the proposed ARA Observatory.

\end{abstract}

\maketitle

\section{Introduction}      



    The observed energy spectrum of comic rays has been extended to beyond \SI{e20}{\eV} \cite{HiRes2009-UHECR-Stereo, Auger2010-EeVSpectrum}, but little is known about their origins and acceleration mechanism, which are important questions in astrophysics \cite{NRC-Report2003}.
    Ultra-high energy cosmic rays (UHECRs) are thought to be of extragalactic origin, such as being produced by active galactic nuclei (AGNs) or gamma-ray bursts (GRBs) \cite{Torres2004}.
    Such UHECRs can generate ultra-high energy ($E>10^{17}$\si{\eV}) neutrinos via photo-pion production or proton-proton interaction:
    \begin{eqnarray*}
        p+\gamma & \rightarrow & \Delta^{+} \rightarrow n+\pi^{+}, \\   
        p + p   & \rightarrow & \pi^{+}\pi^{-}\pi^{0},
    \end{eqnarray*}
    and the subsequent decays of charged pion and muon, e.g.,
    %
    \begin{eqnarray*}
        \pi^{+}\rightarrow \nu_{\mu} + \mu^{+} \rightarrow  \nu_{\mu} + \bar{\nu}_{\mu} + \nu_{e} + e^{+},    
    \end{eqnarray*}
    where the targets can be the intergalactic medium near the astrophysical sources \cite{Stecker1991, *Stecker1992-Erratum, Waxman1997} (see Ref.~\cite{Halzen2002, *Becker2008} for a review), or the cosmic microwave background (CMB) photons (the Greisen-Zatsepin-Kuzmin process) \cite{Greisen1966, *Zatsepin1966, Beresinsky1969}.
    %
    Neutrinos originating from the GZK process are known as the cosmogenic neutrinos (or GZK neutrinos), which are guaranteed to exist based on the fact that both initial-state particles, i.e., the UHECR and the CMB photon, have been observed and that the notion is consistent with the observed GZK cutoff in the cosmic ray spectrum \cite{HiRes2008, *Auger2008}.
    Since the production of ultra high energy cosmic neutrinos (UHECNs) are tightly connected with UHECRs, such neutrino spectrum can help to resolve the puzzles of cosmic rays such as their composition \cite{Hooper2005a, Ahlers2009}, the energy spectrum at the sources, and the cosmological evolution of the sources \cite{Seckel2005, Kotera2010}.


    %
    Besides the overall spectrum, the relative flux ratio between different neutrino flavors, or briefly, the flavor ratio, can also provide information about the physical properties of UHECR sources. For example, the transition of flavor ratio at the source from $f_{e}^{S}:f_{\mu}^{S}:f_{\tau}^{S} = 1:2:0$ (pion source) to $0:1:0$ (muon-damped source) with increasing energies due to synchrotron energy loss of muons can be used to constrain the strength of cosmic magnetic field \cite{Kashti2005, Lipari2007, Huemmer2010} (neutrinos and antineutrinos are counted together because they are hard to be discriminated in the UHE neutrino detection).
    Furthermore, during the propagation from the source to the Earth, the flavor composition of a neutrino would oscillate \cite{Learned1995, *Athar2000}, and may even be altered by some new physics beyond the Standard Model, such as the neutrino decay \cite{Beacom2003a, *Beacom2004a, Barenboim2003, Meloni2007, Maltoni2008, Bhattacharya2010a}, pseudo-Dirac states of neutrinos \cite{Beacom2004}, sterile neutrinos with tiny mass differences \cite{Keranen2003}, the violation of CPT or Lorentz invariance \cite{Barenboim2003, Hooper2005, Bhattacharya2010a}, and the quantum decoherence \cite{Hooper2005, Bhattacharya2010a, Lai2010} (see Ref.~\cite{Pakvasa2008} for a review). With extremely high energies and long traveling distances ($>10$Mpc), the flavor ratio of UHECNs can also help constrain neutrino oscillation parameters \cite{Winter2006, Serpico2006, Xing2006, Majumdar2007, Rodejohann2007, Choubey2008} and probe exotic physics in the parameter regime inaccessible on Earth.
    \footnote{In fact, most of these references consider neutrinos with energy $>$ PeV.}

    To detect UHECNs, enormous amount of matter is required for the target due to their low flux and tiny interaction cross section.
    %
    %
    There are four major detection strategies depending on either neutrinos interact with nucleons via the neutral current (NC) interaction, $ \nu_{l}+N \rightarrow  \nu_{l}+X$, or via the charged current (CC) interaction, $\nu_{l}+N  \rightarrow  l^{-}+X$, where $l$ stands for lepton and $X$ for hadronic debris that will develop into hadronic showers.
    %
    %
    The first approach is to observe the optical Cherenkov lights emitted by secondary charged particles and showers by an array of optical sensors (e.g.~photomultiplier tubes) deployed deep in the medium, e.g.~under-ice arrays such as AMANDA \cite{AMANDA-II2008} and IceCube \cite{Halzen2006, *IceCube2011-EHENuFlux} at the South Pole, and Baikal, ANTARES, NESTOR, NEMO, KM3NET \cite{KM3NeT2006} underwater.
    %
    %
    The second one is to detect horizontal or Earth-skimming neutrino-induced air showers, such as Pierre Auger Observatory \cite{Auger2008-Neutrino, *Auger2009-Neutrino} and HiRes \cite{HiRes2008-Neutrino}.
    %
    The third approach is to detect acoustic waves generated by the showers, which is still in the R\&D stage \cite{Askaryan1957, *SAUND2005}.
    %
    The last and a very promising one is to observe the radio Cherenkov emission from the neutrino-induced showers in dense media through the Askaryan effect \cite{Askaryan1962, *Askaryan1965}. Showers propagating in dense media would develop about \SI{20}{\percent} of excess negative charges and would emit Cherenkov radiation, which is coherent in the radio frequencies up to a few \si{\GHz} due to the compact shower size.
    %
    %
    This effect has been verified in a series of beam experiments \cite{SLAC-Askaryan2001, *SLAC-Askaryan2005, *SLAC-Askaryan2007}.
    The radiated power in the coherent regime is proportional to the square of net charges, which is roughly proportional to the shower energy, making this technique especially sensitive to UHE showers and thus UHECNs.
    Another advantage of this approach is the long radio attenuation length in some natural media, e.g.~the Polar ice and salt, with lengths typically of order of 0.1 to \SI{1}{\km}, and therefore detectors are able to monitor large target volume and achieve greater sensitivity.
    %
    Observatories of this type are: the FORTE satellite \cite{FORTE2004} looking for neutrino signals from the Greenland ice; the balloon-borne antenna array ANITA \cite{ANITA2009, *ANITA-lite2006, *ANITA1-Neutrino, *ANITA2-2010, *ANITA2-2010-erratum} overlooking the Antarctic ice; and radio telescopes looking for signals from the lunar regolith, e.g.~GLUE \cite{GLUE2004} and LUNASKA \cite{LUNASKA2010}.
    %
    There are also attempts to deploy antennas inside the target media in order to lower the threshold energy (to about \SI{e17}{\eV}), e.g.~under-ice antenna arrays RICE \cite{RICE2006}, ARA \cite{ARA2011}, and ARIANNA \cite{ARIANNA2007, *ARIANNA2010} in the Antarctic ice; and SalSA \cite{SalSA2002} in salt dome.





    It is impossible to distinguish between neutrino flavors from the NC interactions because their only products are hadronic showers.
    %
    On the other hand, the charged leptons produced in the CC interactions have different energy-loss characteristics in the medium, which provides an opportunity to identify the flavor.
    Optical Cherenkov neutrino telescope, e.g.~IceCube, is able to identify the flavors by event topologies. The muon from the $\nu_{\mu}$ CC interaction leaves a track; whereas showers from all flavor's NC events and $\nu_{e}$'s CC events lead to localized trigger patterns; and there are double-bang and lollipop events unique to $\nu_{\tau}$ induced by the $\tau$ decays.
    %
    Beacom et al.~\cite{Beacom2003, *Beacom2005} proposed a method to deduce the neutrino flavor ratio from the measured events of different types, and its feasibility has been widely investigated (e.g.~\cite{Serpico2006, Winter2006, Meloni2007, Majumdar2007, Lai2009}).
    %
    However, the instrumented volume, currently of cubic kilometer scale for the largest, limits the event rate for UHE neutrinos, which renders it challenging to distinguish between $\nu_{\mu}$ and $\nu_{\tau}$ events at UHEs as the decay length of $\tau$ lepton exceeds the detector size \cite{Bugaev2004}.


    For radio Cherenkov telescopes such as ARA, the situation is a somewhat different. Though this approach cannot detect the track by a single charged particle, the $\nu_{\mu}$ and $\nu_{\tau}$ CC events can in principle be separated through the amount of energy deposited by leptons into electromagnetic and hadronic showers \cite{Bugaev2004}. It has also been pointed out that different types of showers at UHEs can be distinguished according to their elongation by the Landau-Pomeranchuk-Migdal (LPM) effect  \cite{LPM1953a, *LPM1953b, *LPM1956, AZ1997, *AZ1998}. In addition, $\nu_{e}$ CC events can generate mixed showers of both types and should have its own characteristic signal feature \cite{AVZ1999}.
    The feasibility of this method has been investigated in SalSA \cite{SalSA-Flavor2006}.



    Apart from the event signatures, the direction distribution of neutrino events also manifest themselves in the the energy-loss properties of leptons since neutrinos from different directions propagate through different amounts of matter in the Earth.
    %
    For example, $\nu_{\tau}$ can undergo the regeneration process ($\nu_{\tau} \rightarrow \tau \rightarrow \nu_{\tau}$) without being absorbed due to the $\tau$ decay \cite{Halzen1998, Bugaev2004}, and as a result can exhibit a higher flux in the up-going directions.
    Therefore the neutrino distribution can be a useful tool for measuring the flavor ratio and should be applicable to any detector with sufficient angular resolution of event direction. A similar idea that takes advantage of the event direction distribution has been proposed to constrain neutrino-nucleon cross sections \cite{Connolly2011}.

    In this paper we focus on the cosmogenic neutrinos and consider three expected flavor ratios when they arrive at the Earth's surface, that is, $1:1:1$ expected in the standard oscillation scenario \cite{Learned1995, *Athar2000}; and 6:1:1 as well as 0:1:1 ratios predicted in the neutrino decay models \cite{Beacom2003a} with normal and inverted neutrino mass hierarchy, respectively. The sum of ratios is normalized to unity in the following sections.
    With the detector configuration based on ARA \cite{ARA2011} currently under construction at the South Pole, we demonstrate the feasibility of extracting the flavor ratio from the event direction distribution, while the inference on the flavor ratio at the source is beyond our scope.


      In the next section we present the simulation setup, and derive the expected event direction distribution for each flavor in Section \ref{sec:DD}. The procedures for hypothetical experiments and the extraction of flavor ratios from the direction distribution of pseudo-data are described in Section \ref{sec:GenData} and \ref{sec:FitData}. The successful probability of this method and the flavor ratio resolution as functions of the number of observed events and angular resolution of neutrino direction are reported in Section \ref{sec:ProbSuccess-Resolution}.

\section{Simulation Setup\label{sec:Simulation}}


    Our simulation is built by integrating existing packages that consists of two parts. The first part is the propagation of the neutrinos and the secondary charged leptons.
    The second part is the simulation of event detection, including the conversion of particle energy losses to showers, the development of Cherenkov radiations from showers, the propagation of Cherenkov radiations to detectors, and the calculation of detector responses. The planned configuration of ARA \cite{ARA2011} and the ice properties at the South Pole are adopted as the setting in our simulations.


 \subsection{Neutrino Generation and Propagation Using MMC \label{sec:MMC}}
     The Muon Monte Carlo (MMC) package \cite{*[{}] [{; http://dima.lbl.gov/work/MUONPR/.}] MMC2008} is employed to generate neutrinos and propagate all types of neutrinos and secondary charged leptons. In MMC, interaction cross sections of neutrinos are evaluated based on Ref.~\cite{GQRS1998} with CTEQ6 parton distribution functions \cite{CTEQ6}.

     For charged leptons, energy losses via ionization, pair production, bremsstrahlung, photonuclear interaction, and decay are taken into account.
     The Kelner-Kokoulin-Petrukhin (KKP) \cite{KKP1995, *KKP1997} and Bezrukov-Bugaev (BB) \cite{BB1981} parameterizations are chosen for cross section calculations of bremsstrahlung and photonuclear interaction, respectively. We do not propagate secondary electrons and regard them as losing all of their energies to shower developments within a short distance once they are generated.
     Taus decay into electron, muon, or hadrons are considered, and hence the $\nu_{\tau} \rightarrow \tau \rightarrow \nu_{\tau}$ regeneration in the propagation.
     %

     Monoenergetic neutrinos are generated isotropically at the Earth's surface and start their propagation to the detection volume. The Earth model provided in MMC simulation code is used, where the density is calculated based on the Preliminary Reference Earth Model (PREM)~\cite{PREM1981} while the composition as well as the topography on the Earth's crust are not considered. The detection volume, which is the South Polar ice sheet in the vicinity of ARA, is approximated by a cylindrical ice volume, centered at \SI{1}{km} below the ice surface with a radius of \SI{8}{\km} and a height of \SI{2}{\km}.
     In the propagation, all neutrinos and charged leptons are tracked until they either are absorbed by the Earth or exit the detection volume; energy losses greater than \SI{1}{\peta\eV} are treated stochastically.
     Only those neutrino events traversing the detection volume with at least one stochastic energy loss are reserved for the event detection.
     The cutoff energy is chosen based on the consideration that the radio Cherenkov signals emitted by showers below this energy would not be strong enough to trigger the detector efficiently.

 \subsection{Neutrino Event Detection Using SADE \label{sec:SADE}}

  The Simulation of Askaryan Detection and Events (SADE) package \cite{SADE} is used for simulating neutrino detection. Neutrino events recorded in the previous step, as described in Section \ref{sec:MMC}, are processed individually. Every energy loss exceeding \SI{1}{\peta\eV} within the detection volume is converted into showers of corresponding types. Hadronic products of neutrino CC and NC interactions, $\tau$ decay, and photonuclear interactions are turned into hadronic showers, while secondary electron, pair production, and bremsstrahlung are turned into electromagnetic showers.
  %
  %
  In general, a neutrino event can generate multiple showers in the detection volume. For example, a $\nu_{e}$ event having CC interaction would generate a hadronic shower and an electromagnetic shower, whereas a $\nu_{\mu}$ or a $\nu_{\tau}$ event may have several to even hundreds of showers produced by the secondary $\mu$ or $\tau$ lepton, respectively.

  In SADE, shower characteristics and the frequency spectrum of Cherenkov radiation are calculated with analytic formulae, where the latter is primarily based on Ref.~\cite{AVZ2000} but with a little bit modification of parametrization to account for the decoherence due to the longitudinal and the lateral spreads of shower.

  A ray-tracing routine then finds the paths of both the direct and the reflected (due to the ice-air interface at the surface) rays connecting each shower to each antenna in the ice whose index of refraction varies with depth. The flight time, the radiation spectrum taken into account the frequency-dependent attenuation along the path as well as the polarization at the antenna are calculated for each ray. The spectrum is then Fourier-transformed to the electric field received by the antenna in the time domain accordingly.

\begin{figure}
   \includegraphics[width=0.9 \linewidth]{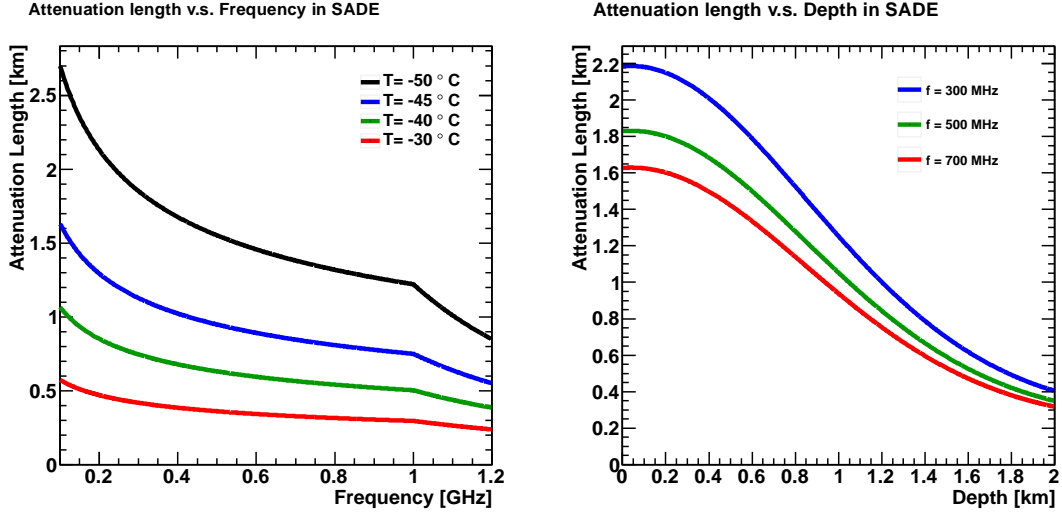}
   \caption{\label{fig:IcePlot}Attenuation length used in SADE \cite{SADE} as a function of the radiation frequency at different temperatures (\SIlist{-30;-40;-45;-50}{\celsius}, left panel) and as a function of the depth at different frequencies (\SIlist{300;500;700}{\MHz}, right panel), respectively.}
 \end{figure}

    Ice properties such as the index of refraction \cite{Kravchenko2004, *Landsman2012}, the temperature \cite{Woschnagg2001}, and the radio attenuation length \cite{Barwick2005, ARA2011}
    are based on the results of \textit{in situ} measurements at the South Pole. The index of refraction $n$ and the ice temperature $T$ (in \si{\degreeCelsius}) depend only on the depth (in \si{\km}), $|z|$,
     \begin{eqnarray}
         n(z)   &   =   &    1.78 - (1.78 -1.35) \exp (-13.2 |z|),\label{eq:IceN}  \\
         T(z)   &   =   &    -51.5 - 0.45319 |z| + 5.822 |z|^{2}.
     \end{eqnarray}
    The attenuation length in turn depends on the ice temperature as well as the radiation frequency, and is plotted in Fig.~\ref{fig:IcePlot}. Note that the rising ice temperature with increasing depth renders the attenuation length shorter and thus suppresses the detectability of signals originating from the bottom part of ice. The birefringence of South Polar ice, which is the polarization dependence of the wave speed and the attenuation due to the crystal anisotropy and orientation of ice, is not considered here. It is reported \cite{Kravchenko2011} that the birefringence is observed at the bottom half of the ice sheet and will reduce about \SI{5}{\percent} of the neutrino detection volume.

    The detector based on the planned configuration of ARA \cite{ARA2011} is a hexagonal array of $37$ antenna stations arranged in a triangular grid with \SI{2}{\km} spacing. The array covers a total area of about 100 km$^2$.
    Each station is an autonomously operating cluster of eight vertically polarized (Vpol) and eight horizontally polarized (Hpol) antennas evenly deployed on four vertical strings with each string placed on one vertex of a square. Each string is at a maximum depth of \SI{200}{\m} and is loaded with two antenna pairs, where each antenna pair contains a Vpol antenna and an Hpol antenna.
    The values of parameters for the detector settings are listed in Table \ref{table:SADE-Parameter}.

    Given the incident electric field $\vec{E}$ at an antenna, the received signal voltage of the antenna $V_{\textrm{signal}}$ is
    \begin{eqnarray}
        V_{\textrm{signal}} &=& \frac{1}{2} \vec{E} \cdot \vec{h}_{\textrm{eff}},
    \end{eqnarray}
    with
    \begin{eqnarray}
        |\vec{h}_{\textrm{eff}}| &=& 2 \sqrt{ \frac{A_{\textrm{eff}}Z_{\textrm{ant}} }{n Z_{0}} }, \\
        A_{\textrm{eff}} &=& \frac{Gc^{2}}{4\pi f^{2}},
    \end{eqnarray}
where $\vec{h}_{\textrm{eff}}$ and $A_{\textrm{eff}}$ are the effective height and the effective area of the antenna, respectively; $Z_{\textrm{ant}}$ the antenna impedance, $Z_{0} \simeq 377 \Omega$ the impedance of free space, $n$ the index of refraction of surrounding medium, $G$ the antenna gain, and $c$ the speed of light in vacuum. The direction of effective height for Vpol antennas is in the vertical direction $\hat{z}$, while it is in the azimuthal direction $\hat{\phi}$ for the Hpol. In the simulation, the frequency response of the antenna is assumed to be a single perfect passband, and the effective height is approximated by a single value evaluated at the central frequency of the passband. For each antenna, signals coming from different showers are summed in the time domain. We neglect the contribution from the reflected signals because they would have sufficient time delays and would suffer more attenuation than the direct ones due to longer path length and are more difficult to be reconstructed after passing through the less compact snow near the surface (firn) where the index of refraction changes rapidly (see Eqn.~\ref{eq:IceN}).

     The root mean square (RMS) thermal noise voltage of an antenna $V_{\textrm{rms}}$ is defined as
     \begin{equation}
        V_{\textrm{rms}} = \sqrt{k_{\textrm{B}} T_{\textrm{sys}} Z_{\textrm{ant}} B },
     \end{equation}
     where $k_{\textrm{B}}$ is Boltzmann's constant, $T_{\textrm{sys}}$ the system noise temperature of the receiving antenna system, $f$ the radiation frequency and $B$ the frequency bandwidth of the antenna. The values of antenna parameters used in the simulation are summarized in Table~\ref{table:SADE-Parameter}.


     The trigger conditions for a detected event require that
     \begin{inparaenum}[\itshape i\upshape)] 
     \item  the received voltage of a triggered antenna should exceed three times of its RMS noise voltage (i.e., $V_{\textrm{signal}} \geq 3V_{\textrm{rms}}$);\\
     \item  at least five out of sixteen antennas in a station are triggered; and\\
     \item  at least one station is triggered.
     \end{inparaenum}

   \begin{table}[h]
    \caption{\label{table:SADE-Parameter}
       Parameters for the detector configuration and the antennas, and their values used in the simulation.
    }
    \begin{ruledtabular}
    \begin{tabular}{@{\extracolsep{\fill}}lr}
    Parameter (unit) & Value \\ \hline
    Station spacing (km)    &   2   \\  
    Radius of string (m)          &   10\\      
    Number of strings per station   &   4\\
    Separation between paired antennas (m)  &   5\\
    Vertical spacing between antenna pairs (m) & 20\\
    Maximum antenna depth (m)     &   200 \\
    Number of Vpol antennas per station      &  8\\
    Number of Hpol antennas per station      &  8\\
    Vertical antenna configuration  &   Vpol, Hpol above Vpol, Hpol\\
        &      \\
    Vpol frequency band: $B_{\textrm{V}}$ (\si{\MHz})  &  150-850\\
    Hpol frequency band: $B_{\textrm{H}}$ (\si{\MHz})  &  200-850\\
    Antenna impedance: $Z_{\textrm{ant}}$ (\si{\ohm})   &   50  \\
    Antenna gain: $G$   &   1.64\\
    Effective height of Vpol antenna: $|\vec{h}_{\textrm{eff,V}}|$ (cm)    &  11.8\\
    Effective height of Hpol antenna: $|\vec{h}_{\textrm{eff,H}}|$ (cm)    &  11.3\\
    System noise temperature of antenna: $T_{\textrm{sys}}$(K)   &   300\\
    RMS noise voltage of Vpol: $V_{\textrm{rms,V}}$ (V)      &    $1.20\times10^{5}$\\
    RMS noise voltage of Hpol: $V_{\textrm{rms,H}}$ (V)      &    $1.16\times10^{-5}$\\
    Antenna trigger threshold ($V_{\textrm{rms}}$)     &   3\\
    Station trigger threshold (antennas)    &   5\\
    \end{tabular}
    \end{ruledtabular}
  \end{table}

\section{ Angular Distribution of Neutrino Events\label{sec:DD}}



    In the simulation, monoenergetic neutrinos with initial energies $\log_{10}(E_{\nu}/\textrm{eV})=$ \numlist{17;17.5;18;18.5;19;19.5} are generated separately.
   %
   To acquire the expected direction distribution of detected neutrino events with initial energy $E_{\nu}$ and flavor $\alpha$, $D_{\alpha}( E_{\nu},\cos \theta)$, the following information are needed:

    \begin{inparaenum}[\itshape a\upshape)] 
        \item   the flux of isotropic cosmogenic neutrinos for all flavors, $\Phi_{\nu}(E_{\nu})$, as well as the incident flux ratio among three flavors at the Earth's surface, $f_{e}^{E}: f_{\mu}^{E}:f_{\tau}^{E}$, with $\sum_{\alpha} f_{\alpha}^{E} = 1$;

        \item   the interaction probability $P_{\textrm{int}, \alpha}(E_{\nu}, \cos \theta)$, which is defined as the probability that a neutrino or its secondary lepton traverses the Earth in the zenith direction $\cos \theta$ without being stopped and interacts (with at least one energy loss exceeding \SI{1}{\peta\eV}) inside the detection volume, to account for the propagation effect; and

        \item   the detection efficiency to the subsequent shower(s) generated by the neutrino event interacting inside the detection volume, $\epsilon_{ \textrm{det},\alpha}(E_{\nu}, \cos \theta)$.
    \end{inparaenum}
    That is,
   \begin{eqnarray}
        D_{\alpha}( E_{\nu},\cos \theta) &=& \frac{1}{\mathbb{N}}f_{\alpha}^{E} \Phi_{\nu}(E_{\nu}) P_{\textrm{int}, \alpha}(E_{\nu}, \cos \theta) \\ \nonumber
        && \times  \epsilon_{ \textrm{det},\alpha}(E_{\nu}, \cos \theta),
    \label{eq:DirectionDistribution}
   \end{eqnarray}
   with the normalization factor
   \begin{eqnarray}
     \mathbb{N} &=& \sum_{\alpha} \int dE_{\nu} \int_{-1}^{1} d\cos\theta f_{\alpha}^{E} \Phi_{\nu}(E_{\nu})   \\ \nonumber
     && \times  P_{\textrm{int}, \alpha}(E_{\nu}, \cos \theta) \epsilon_{ \textrm{det},\alpha}(E_{\nu}, \cos \theta),
   \end{eqnarray}
   where the distribution has been normalized as a probability distribution independent of the total event rate, $\alpha=e,\mu,\tau$, the superscript $E$ indicates quantities on the Earth, and $\theta$ is the angle between the local vertical axis of the detector ($\hat{z}$) and the direction of neutrino momentum. Hence, events with negative $\cos \theta$ are down-going, whereas those with positive values are up-going. The neutrino energy range considered in this article is $\log_{10}(E_{\nu}/\textrm{eV})= 16.75$--$19.75$.

    Note that more precisely defined interaction probability and detection efficiency should depend not only on the initial neutrino energy but also the amount of energy loss in the detection volume. But since in this article we focus on the direction of events and the angular distribution is insensitive to the amount of energy loss in the detection volume, the calculation of interaction probability $P_{\textrm{int}}(E_{\nu}, \cos \theta)$ and detection efficiency $\epsilon_{\textrm{det}}(E_{\nu}, \cos \theta)$ has averaged over all events with energy loss above the threshold \SI{1}{\peta\eV} in the detection volume. In addition, in our simulation result, for more than about $95\%$ of $\nu_{\mu}$ and $\nu_{\tau}$ events only one shower can be detected, so we did not separate the detection efficiency and the angular distribution into single cascade channel from CC/NC interaction and multiple cascade channel from $\mu$/$\tau$ leptons.

    \subsection{Flux and Flavor Ratio on Earth}

\begin{figure}
        \includegraphics[width=0.7\linewidth]{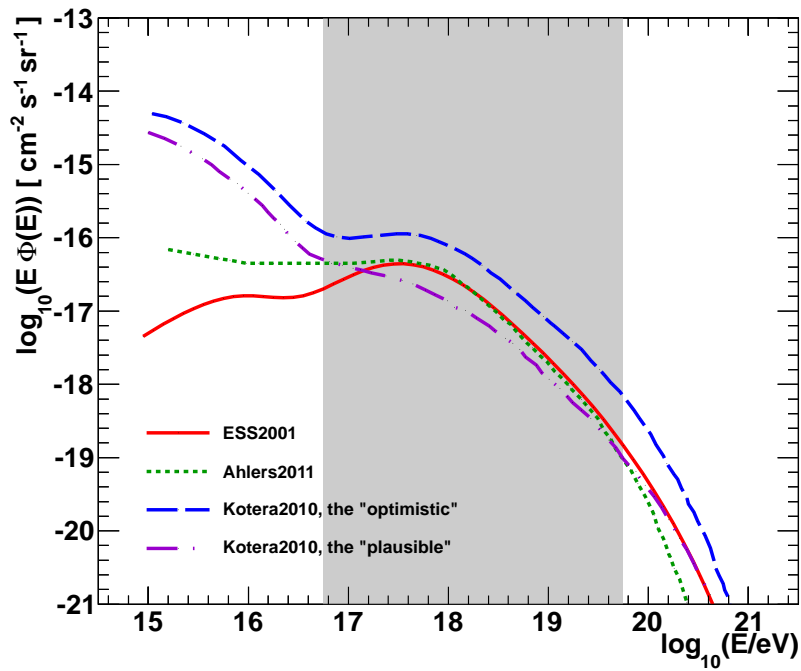}  
        \caption{\label{fig:GZKFlux}
            Differential cosmogenic neutrino fluxes predicted by \cite{ESS2001} (ESS, red solid curve), \cite{Ahlers2010} (green dotted), the optimistic (blue dashed) and the plausible (purple dot-dashed) scenarios in \cite{Kotera2010}. The shaded region indicates the neutrino energy range considered in the analysis, $\log_{10}(E_{\nu}/\textrm{eV})= 16.75$--$19.75$.
        }
    \end{figure}

The cosmogenic neutrino fluxes for all flavors, $\Phi_{\nu}(E_{\nu})$, have been theoretically
predicted in Refs.~\cite{ESS2001, Ahlers2010, Kotera2010} (see Fig.~\ref{fig:GZKFlux}).
    In the following analysis, the neutrino flux from Ref.~\cite{ESS2001} (red solid curve; hereafter, ESS) is assumed. Note that it is the spectral shape that affects the event direction distribution instead of the overall flux normalization.
    Therefore one expects that neutrino fluxes predicted in \cite{ESS2001}, \cite{Ahlers2010} (green dashed curve), and the optimistic scenario in \cite{Kotera2010} (purple dashed curve) should yield similar distributions. The flux predicted in the plausible scenario in \cite{Kotera2010} (blue dashed curve) has a steeper spectrum than others, and we will present its results later in Sec.~\ref{sec:Obs}.

    The flavor ratio of neutrinos arriving at the Earth's surface adopted in our analysis is $f_{e}^{E}: f_{\mu}^{E}:f_{\tau}^{E} = 1/3:1/3:1/3$, as expected in the standard oscillation scenario \cite{Learned1995, *Athar2000}; $0.75:0.125:0.125$ and $0:0.5:0.5$ predicted in the neutrino decay scenarios with normal and inverted neutrino mass hierarchy, respectively \cite{Beacom2003a}. Throughout this paper these flavor ratios are assumed to be energy-independent over the considered neutrino energy range.



    \subsection{Interaction Probability}
\begin{figure}
        \includegraphics[width=0.7\linewidth]{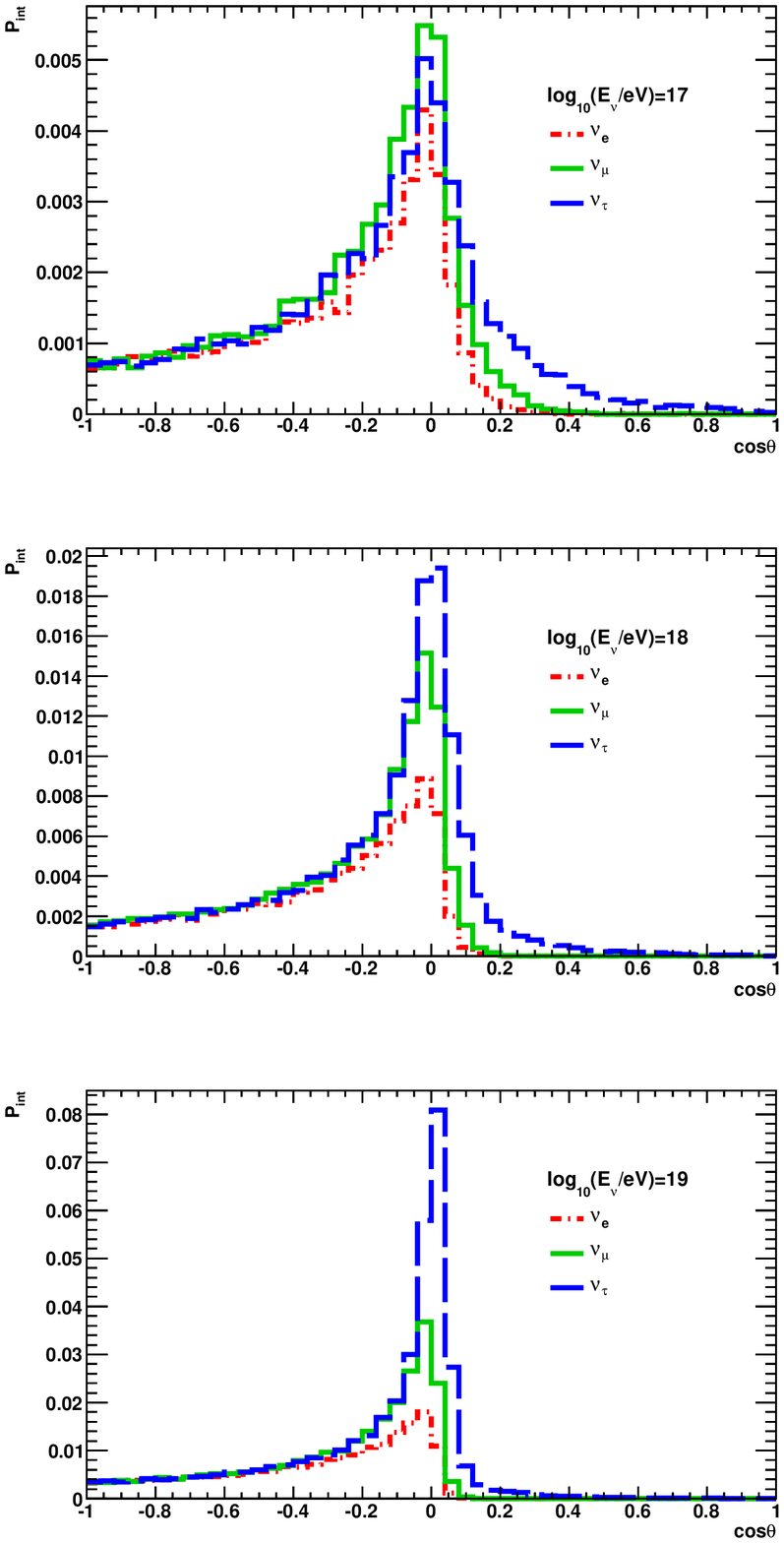}
     \caption{       \label{fig:NuIntProb}
              Interaction probability $P_{\textrm{int}}$ (see text for definition) as a function of the neutrino zenith direction $\cos \theta$ for $\nu_{e}$ (red dash-dotted), $\nu_{\mu}$ (green solid), and $\nu_{\tau}$ (blue dashed), and for initial energies $E_{\nu}=$ \SIlist{e17;e18;e19}{\eV} (from top to bottom panel).
              %
     }
    \end{figure}

    To obtain the interaction probability $P_{\textrm{int}}$ from the simulation results, we first divide the zenith angle of neutrinos $\cos\theta$ into 100 bins with a width of $0.02$. The probability at each bin is defined as the ratio of the number of survival events inside the detection volume to the number of initial incoming events at the Earth's surface.
    %
    %
    The probability does not depend on the azimuthal angle of the neutrino due to the axial symmetry of our Earth model and detection volume.
    %
    %
    The probabilities $P_{\textrm{int}, \alpha}(E_{\nu}, \cos \theta)$ for initial neutrino energies $E_{\nu}=$ \SIlist{e17;e18;e19}{\eV} are shown in Fig.~\ref{fig:NuIntProb}.


    For different flavors of neutrinos with the same initial energy, the interaction probabilities are about the same as $\cos \theta$ approaches $-1$ where neutrinos impinging directly downward into the detection volume, whose size is much smaller than the neutrino interaction length (about several hundred \si{\km} \cite{GQRS1996}).
    Thus the probability is approximately equal to the size of the detection volume divided by the neutrino interaction length for energy transfers greater than \SI{1}{\peta\eV}. The probability for neutrinos impinging downward increases with neutrino energy as the neutrino interaction length decreases.

    The probabilities then increase with $\cos\theta$ and reach a maximum near the horizontal direction ($\cos \theta \simeq 0)$, where the traveling distance of neutrinos becomes comparable to the interaction length and the detection volume has its maximum span. The probabilities for different flavors diverge due to the difference in the energy-loss property between different flavors of charged leptons produced in the CC interactions. The longer the lepton can propagate, the higher the probability is. Contrary to electrons, which would be stopped immediately after their creation and would develop into electromagnetic showers, \si{\exa\eV} $\mu$ and $\tau$ leptons can on the average propagate distance of order of \SI{10}{\km} before come to a stop \cite{Dutta2001}. Muons lose their energy mostly via pair production, while $\tau$ leptons via pair production as well as photonuclear interaction. So the probabilities of finding $\nu_{\mu}$ and $\nu_{\tau}$ are higher than that for $\nu_{e}$. The interaction probability in these directions also increases with neutrino energies because of the decrease of the neutrino interaction length and the lepton propagation range.


    The probability for up-going neutrinos ($\cos\theta>0$) is suppressed as the neutrino traveling distance becomes longer than the interaction length. The higher the initial neutrino energy, the shorter the interaction length, and hence the distribution terminates at smaller $\cos\theta$.
    The Earth attenuates the neutrino both in energy through NC and CC interactions and in number through the stoppage of the charged lepton produced in the CC interaction. As a special case, $\tau$ leptons, having a decay length of about $ 50 \times (E_{\tau}/ \textrm{\si{\peta\eV}} )$ \si{\m}, can transform to $\nu_{\tau}$ through decay before losing too much energy \cite{Halzen1998, Bugaev2004}. Therefore $\nu_{\tau}$ coming from below the horizon would have an apparent larger probability than other two flavors. This is a critical feature for the flavor ratio determination proposed in this article, as we will further ellaborate below.

    \subsection{Detection Efficiency}

    The detection efficiency $\epsilon_{\textrm{det}}$ defined here depends not only on the nature of Cherenkov radiation, the ice properties and the detector configuration, but also on how neutrinos and secondary leptons deposit their energies in the shower. But for the simplicity of computation, we do not decompose it further into a product of the probability that a neutrino or a secondary lepton in specific direction generates shower(s) of specific energy at specific position, times the detection efficiency to each individual shower.

\begin{figure}
         \includegraphics[width=0.7\linewidth]{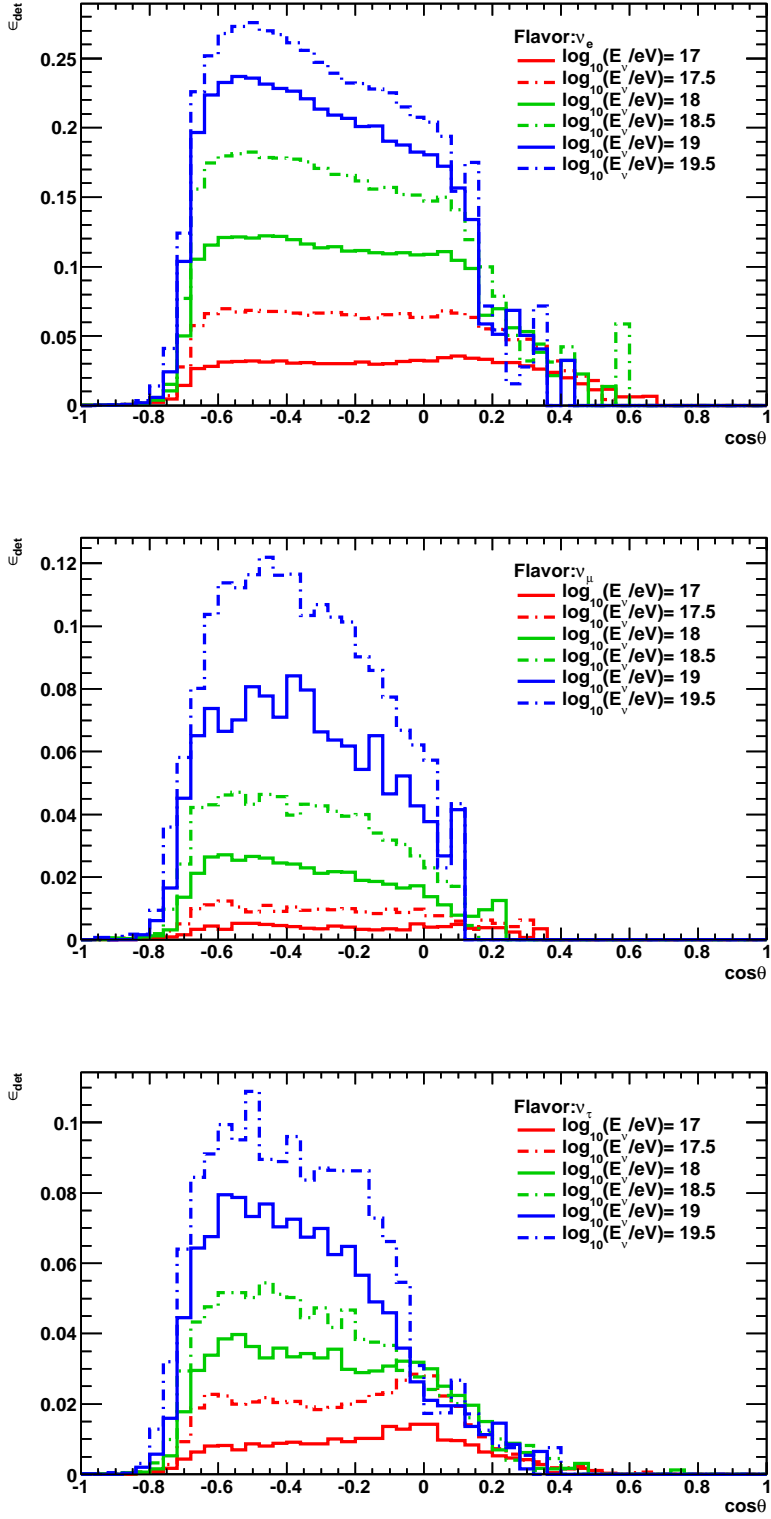}
     \caption{  \label{fig:DetectionEff-2km}  
        The detection efficiency $\epsilon_{\textrm{det}}$ (see text for definition) as a function of neutrino zenithal direction $\cos\theta$ for initial energies $\log_{10}( E_{\nu}/\textrm{eV})=$ 17 (red solid line), 17.5 (red dashed), 18 (green solid), 18.5 (green dashed), 19 (blue solid), 19.5 (blue dashed), and for flavor $\nu_{e}$, $\nu_{\mu}$, and $\nu_{\tau}$ (from top to bottom panel, respectively).
        %
    }
   \end{figure}

    Similar to the definition of the interaction probability, the detection
efficiency is defined as the number of events detected divided by the number
of events interacting inside the detection volume.
    %
    %
    The detection efficiency is hence an averaged quantity over the azimuthal direction and the shower position. The results for different initial neutrino energies and flavors are shown in Fig.~\ref{fig:DetectionEff-2km}.


    The efficiency increases with neutrino energy simply because signal strength is proportional to neutrino energy.
    The Cherenkov cone generated by neutrinos events with $\cos \theta \simeq -0.7$ travels downward and has an opening angle ~\SI{56}{\degree} in ice, so it is less possible to cover the area where the antennas are deployed. This leads to the common cutoff at $\cos \theta \simeq -0.7$ for all neutrino energies.
    In Fig.~\ref{fig:DetectionEff-2km}, we see the fluctuations of the efficiencies for $\nu_{e}$ (top panel) and $\nu_{\mu}$ (middle panel) at $\cos\theta \gtrsim 0.1$. This is due to the smallness of the number of events arriving at the detection volume. Such results are therefore not reliable. However one generic feature remains valid; that is, the up-going neutrino events diminish since their energies are severely damped by the Earth.


    The detection efficiency for $\nu_{e}$ is the highest among the three flavors, because once CC interaction occurs all the neutrino energy is released into showers and strong signals are emitted. For $\nu_{\mu}$, although there are plenty of electromagnetic showers produced by muon, these showers tend to have lower energies so that they are less likely to be detected by the sparse antenna array. This leads to lower detection efficiency of $\nu_{\mu}$, where the NC-induced hadronic showers account for about \SI{80}{\percent} of detected events for \SI{1}{\exa\eV} $\nu_{\mu}$. The situation is similar for $\nu_{\tau}$, where most hadronic showers from photonuclear interaction and electromagnetic showers from pair production do not trigger detector efficiently while hadronic showers induced by NC interaction and $\tau$ decay ($\tau \rightarrow \nu_{\tau}+ \textrm{hadrons}$) account for the most detected events. The convergence of efficiency for up-going $\nu_{\tau}$s of different initial energies results from the degradation of neutrino energy to few \si{\peta\eV} by the regeneration process and NC interaction \cite{Bugaev2004}.


    \subsection{Event Direction Distribution and All-Sky Flavor Ratios of Events \label{sec:DD2}}

\begin{figure}
        \includegraphics[width=0.7\linewidth]{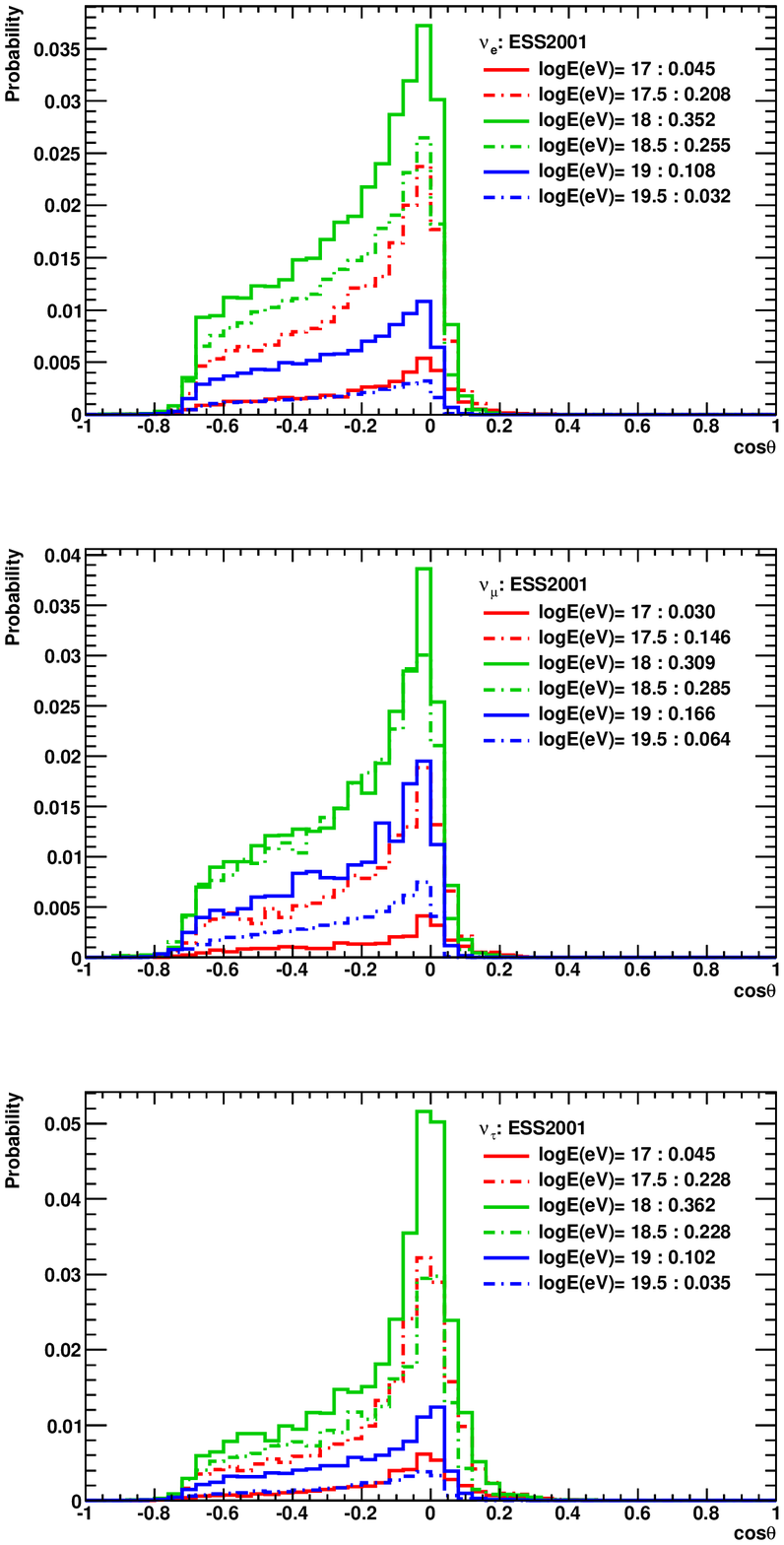}
     \caption{ \label{fig:DD-2km-ESS-AllE} 
        The expected direction distribution integrated over the energy bin, $\log_{10}(E_{j}) \pm 0.25$, of initial neutrinos, where $\log_{10}( E_{j}/\textrm{eV})= $ 17 (red solid), 17.5 (red dashed), 18 (green solid), 18.5 (green dashed), 19 (blue solid), 19.5 (blue dashed), for flavor $\nu_{e}$, $\nu_{\mu}$, and $\nu_{\tau}$ (from top to bottom panel, respectively). The distributions have been normalized so that each one represents its fractional contribution to the corresponding flavor, and the relative fraction of each energy bin is listed in the legend. The ESS neutrino spectrum \cite{ESS2001} is assumed.
    }
   \end{figure}

    Because the initial neutrino energy is sampled only with discrete values of equal logarithmic interval $\Delta \equiv \Delta \log_{10}E_{\nu}=0.5$ in the calculation of $P_{\textrm{int}}$ and $\epsilon_{ \textrm{det}}$, the direction distribution integrated over the $j$-th energy bin ranging from $\log_{10}E_{j}- \Delta/2$ to $\log_{10}E_{j}+ \Delta/2$, is approximated by
    %
    %
    \begin{eqnarray}
       \int D_{\alpha}( E_{\nu},\cos \theta) dE_{\nu} &\simeq &  \frac{1}{\mathbb{N}} f_{\alpha}^{E} [\int_{ E_{j} \times 10^{-\Delta/2} }^{ E_{j} \times 10^{\Delta/2} }\Phi_{\nu}(E_{\nu}) dE_{\nu}]   \\  \nonumber
         & & \times P_{\textrm{int}, \alpha}(E_{j}, \cos \theta) \epsilon_{ \textrm{det},\alpha}(E_{j}, \cos \theta),
    \label{eq:DirectionDistribution}
    \end{eqnarray}
    with $\log_{10}(E_{j}/\textrm{eV})=$ \numlist{17;17.5;18;18.5;19;19.5}. The results assuming the ESS neutrino flux are plotted in Fig.~\ref{fig:DD-2km-ESS-AllE}, where the total area under the distributions for each flavor has been normalized to unity and the relative fraction contributed by each bin is also shown. It appears that \si{\exa\eV} neutrinos contribute the most to the detected events for every flavor, because of the compromise between two competing effects: the decrease in the neutrino flux versus the increase in the detection efficiency with neutrino energy.

\begin{figure}
        \includegraphics[width=0.7\linewidth]{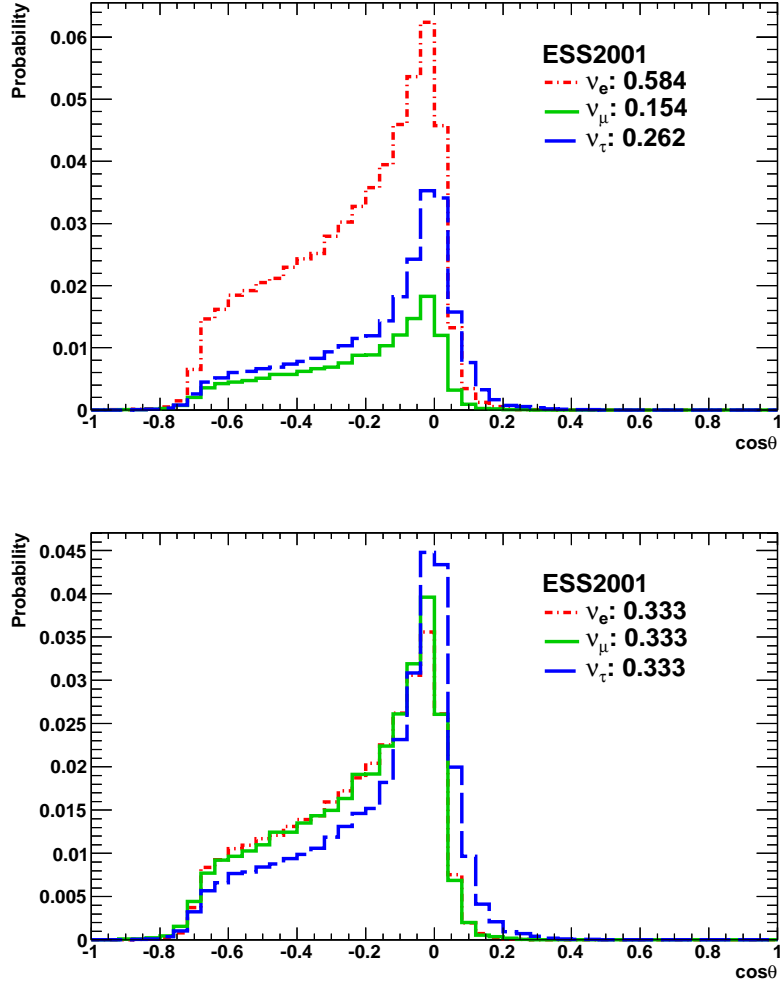}
     \caption{ \label{fig:DD-2km-Sum-ESS} 
        Top panel: the expected direction distribution of detected neutrino events for each flavor $\alpha$, $D_{\alpha}(\cos\theta)$, assuming the ESS flux \cite{ESS2001} and the incident flux ratios $f_{e}^{E}:f_{\mu}^{E}:f_{\tau}^{E} = 1/3:1/3:1/3$. The all-sky flavor ratio of detected events $f_{e}:f_{\mu}:f_{\tau}$ is $0.584:0.153:0.263$.
        Bottom panel: the expected direction distributions are scaled to equal fraction, $1/3$ for each flavor, for comparing the curve shapes.
     }
    \end{figure}

    Finally, after summing over all distributions of different energy bins, the expected event direction distribution for each flavor $D_{\alpha}(\cos\theta)$ is obtained, which is shown in Fig.~\ref{fig:DD-2km-Sum-ESS}.
    We define the ``all-sky'' flavor ratio of detected events, $f_{e}:f_{\mu}:f_{\tau}$, as the event ratios among flavors after integrating over all zenith directions. If the ESS neutrino flux and incident flavor ratios at the Earth's surface, $f_{e}^{E}:f_{\mu}^{E}:f_{\tau}^{E} = 1/3:1/3:1/3$, are assumed, we find that $f_{e}:f_{\mu}:f_{\tau} = 0.584:0.154:0.262$.
    The $\nu_e$ events account for the most portion because of their higher detection efficiency. The $\nu_{\tau}$ events have a different shape compared to the other two flavors especially in the horizontal and up-going directions, primarily due to its special interaction probability (see the graph in the right panel, Fig.~\ref{fig:DD-2km-Sum-ESS}).
    These will be used to extract the flavor composition at the Earth's surface ($f^{E}$s) in the next section. However, the resemblance between $\nu_{e}$ and $\nu_{\mu}$ distributions will lead to the degeneracy in the flavor ratio extraction, and an extra constraint is required, for example, the $\nu_{\mu}$-$\nu_{\tau}$ symmetry.



\begin{figure}
 \begin{center}
  \includegraphics[width=0.6\linewidth]{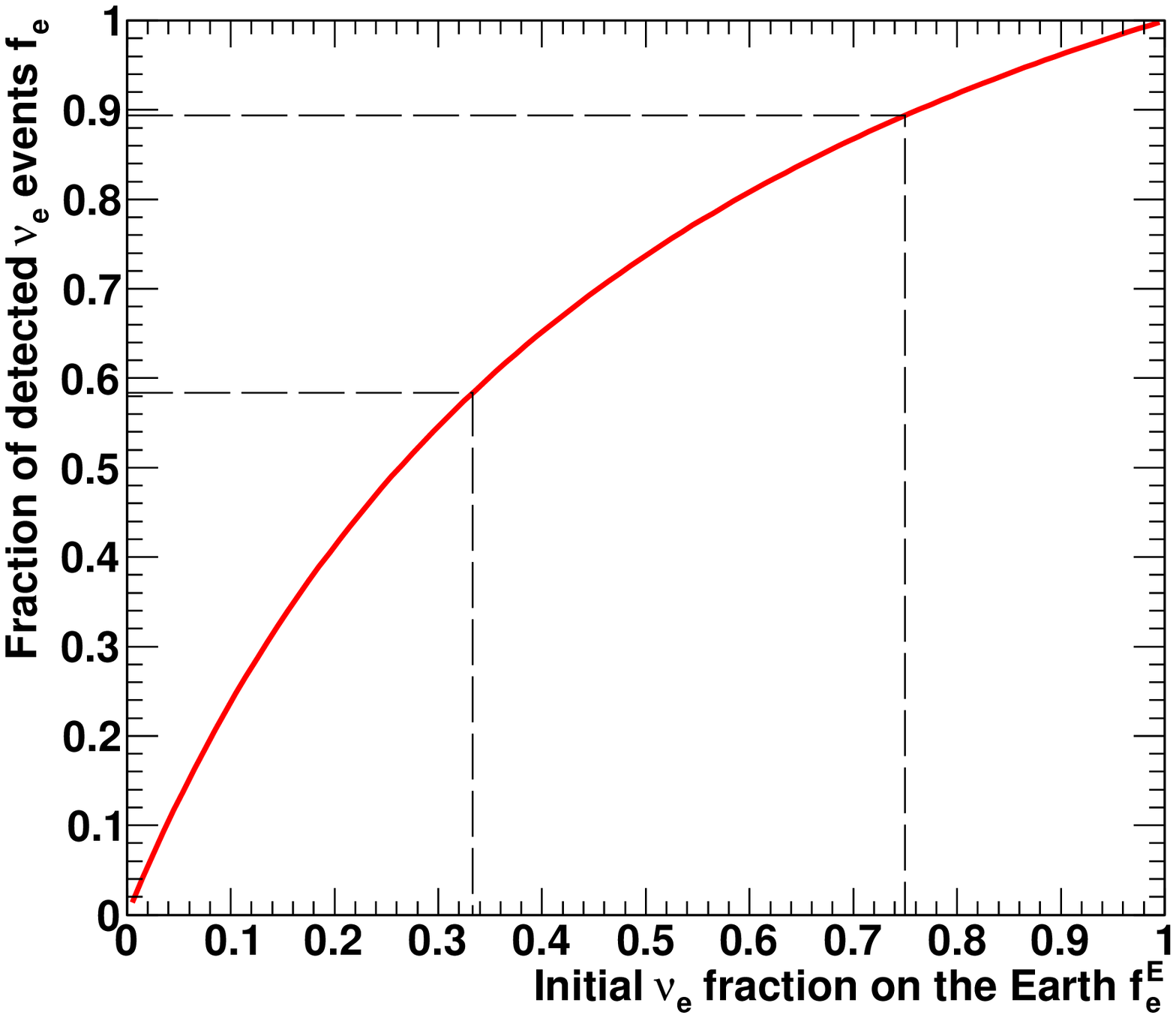}    
 \end{center}
  \caption{ \label{fig:eFlavorConversion-2km}
        The relation of the fraction of detected $\nu_{e}$ events, $f_{e}$ versus the $\nu_{e}$ fraction of incident neutrino flux at the Earth's surface, $f_{e}^{E}$, where the ESS flux model \cite{ESS2001} is assumed.
        The values for the standard case ($f_{e}^{E}:f_{\mu}^{E}:f_{\tau}^{E} = 1/3:1/3:1/3$) and for the neutrino decay case ($0.75:0.125:0.125$) predicted in \cite{Beacom2003a} are labeled.
    }
 \end{figure}

    The event ratio $f$ for other initial flavor ratio $f^{E}$ can be derived from the result above, which is denoted by $f_{0}$ and $f_{0}^{E}$. The $\nu_{e}$ event ratio is
    \begin{equation}
      f_{e} = \frac{ f_{e,0} f_{e}^{E} f_{\mu,0}^{E} f_{\tau,0}^{E}  }{ f_{e,0} f_{e}^{E} f_{\mu,0}^{E} f_{\tau,0}^{E} + f_{\mu,0} f_{e,0}^{E} f_{\mu}^{E} f_{\tau,0}^{E} + f_{\tau,0} f_{e,0}^{E} f_{\mu,0}^{E} f_{\tau}^{E} },
      \label{eq:RatioConversion}
    \end{equation}
    and similarly for $f_{\mu}$ and $f_{\tau}$. The conversion from $f$s to $f^{E}$s can be done by just interchanging $f$ with $f^{E}$. The relation between $\nu_{e}$ event ratio $f_{e}$ and initial flavor ratio $f_{e}^{E}$ is plotted in Fig.~\ref{fig:eFlavorConversion-2km}, assuming ESS neutrino flux.

\section{Pseudo-Observation and Flavor Ratio Extraction\label{sec:Obs}}



    To investigate the discriminating power of flavor ratio reconstruction using event direction distribution, we generate pseudo-observation data from simulated events and fit the direction distribution to extract the flavor ratios, and repeat the processes to determine the statistical uncertainty of the extracted ratio. Results assuming different incident flux ratios, numbers of observed events, and angular resolution of detector are then presented.

    \subsection{Pseudo-Data Samples \label{sec:GenData}}


    The pseudo-data sample is prepared in three steps.
    First, neutrino events with different flavors and directions are randomly generated according to the expected angular distributions, $D_{\alpha}(\cos\theta)$.
    Secondly, the zenith angle, $\theta$, of each pick-up event is smeared by adding a Gaussian distributed random number with zero mean and the standard deviation $\Delta \theta$ equal to an assigned experimental angular error in reconstructed neutrino direction.
    Although multiple cascade events in principle have different angular resolution than single cascade ones, but for their rareness we just assigned the same resolution for both types of events.
    Finally, this event collection is evenly divided into subsets, where each data set represents a hypothetical experimental data sample with a total number of detected events equal to $N_{\textrm{obs}}$ and these data sets constitute a statistical ensemble.

    In the following, $N_{\textrm{obs}}$ varies from $50$ to $500$ with an increment of $50$, while $\Delta\theta$ from \SI{0}{\degree} to \SI{6}{\degree}.

    \subsection{Fitting Pseudo-Data \label{sec:FitData}}

    To construct the fitting function for the hypothetical experimental data samples with angular resolution $\Delta \theta$, the expected direction distribution for each flavor $\alpha$ is convolved with a Gaussian resolution function of standard deviation $\Delta \theta$ in $\theta$ space, $G(\theta' ; \theta, (\Delta\theta)^{2})$. The convolved distribution,
    \begin{equation}
        M_{\alpha}(\theta ; \Delta\theta) \propto \int  D_{\alpha}(\theta) G(\theta '; \theta, (\Delta\theta)^{2}) d\theta ' ,
    \end{equation}
    is then normalized so that
    \[
        \int_{-1}^{1} M_{\alpha}(\cos\theta; \Delta\theta)d\cos\theta = 1,
    \]
    and the probability density function of event direction can be expressed as
    \begin{eqnarray}
        P(\cos \theta ; f_{1},f_{2}, \Delta\theta ) &=& f_{1}M_{e}(\cos \theta) + (1-f_{1})   \\ \nonumber
        & & \times [ f_{2} M_{\mu}(\cos \theta) +(1-f_{2}) M_{\tau}(\cos \theta)],
    \end{eqnarray}
    where $f_{1}$ and $f_{2}$ are unknown fraction coefficients with values between zero and one. This expression ensures $f_{1}$ and $f_{2}$ are independent of each other, and they are simply related to the event flavor ratios by $f_{e} = f_{1}$, $f_{\mu} = (1-f_{1})f_{2}$, and $f_{\tau} = (1-f_{1})(1-f_{2})$.

    The maximum likelihood estimation is applied for flavor ratio extraction. For an experimental data set with total events of $N_{\textrm{obs}}$, the likelihood function is defined as
    %
    \begin{eqnarray}
        L( \cos \theta_{i}; f_{1}, f_{2}, \Delta\theta) & = &   \prod_{i=1}^{N_{\textrm{obs}}} P(\cos \theta_{i} ; f_{1},f_{2}, \Delta\theta ), 
    \end{eqnarray}
    where the subscript $i=1,2,\ldots,N_{\textrm{obs}}$ stands for the $i$-th event. The true values of $f_1$ and $f_2$ are estimated by maximizing $L$, or equivalently minimize the negative log-likelihood (NLL), 
    %
    \begin{eqnarray}
      -\ln L  & = &   -\sum_{i=1}^{N_{\textrm{obs}}} \ln P(\cos \theta_{i} ; f_{1},f_{2}, \Delta\theta ).
    \end{eqnarray}
    We perform grid search in the parameter space to find the minimum of NLL and the associated best fit values of $f_1$ and $f_2$, denoted as $\hat{f}_{1}$ (or $\hat{f}_{e})$ and $\hat{f}_{2}$. A reconstruction is defined as failed if $\hat{f}_{1}$ or $\hat{f}_{2}$ is outside either boundary (zero or one). It may result from the insufficient statistics of events, or that the true value is near the boundary.
    \footnote{When doing the fitting, $\hat{f}_{1}$ and $\hat{f}_{2}$ are permitted to have unphysical values.}



  However, an extra constraint is required to avoid multiple solutions arising from the similarity in distribution shape between $\nu_{e}$ and $\nu_{\mu}$, as pointed out in Section \ref{sec:DD2}. The $\nu_{\mu}$-$\nu_{\tau}$ symmetry, i.e.~the incident $\nu_{\mu}$ and $\nu_{\tau}$ fluxes on the Earth are of equal amount ($f_{\mu}^{E} = f_{\tau}^{E}$), is imposed for this purpose. As a consequence, the value of $f_2$ is known and fixed, and there is only one parameter $f_1$, the fraction of $\nu_{e}$ events, left to be fitted. In addition, for simplicity, we assume the observer knows the exact shape of neutrino spectrum; that is, the neutrino fluxes assumed for generating pseudo-data samples and fitting function are identical. The latter constraint will be relaxed in the next section and the mismatch in shape between real and expected spectra will introduce a systematic bias to the extracted ratio.

    \subsection{Successful Probability and Flavor Ratio Resolution \label{sec:ProbSuccess-Resolution}}

\begin{figure*}
         \includegraphics[width=0.7\linewidth]{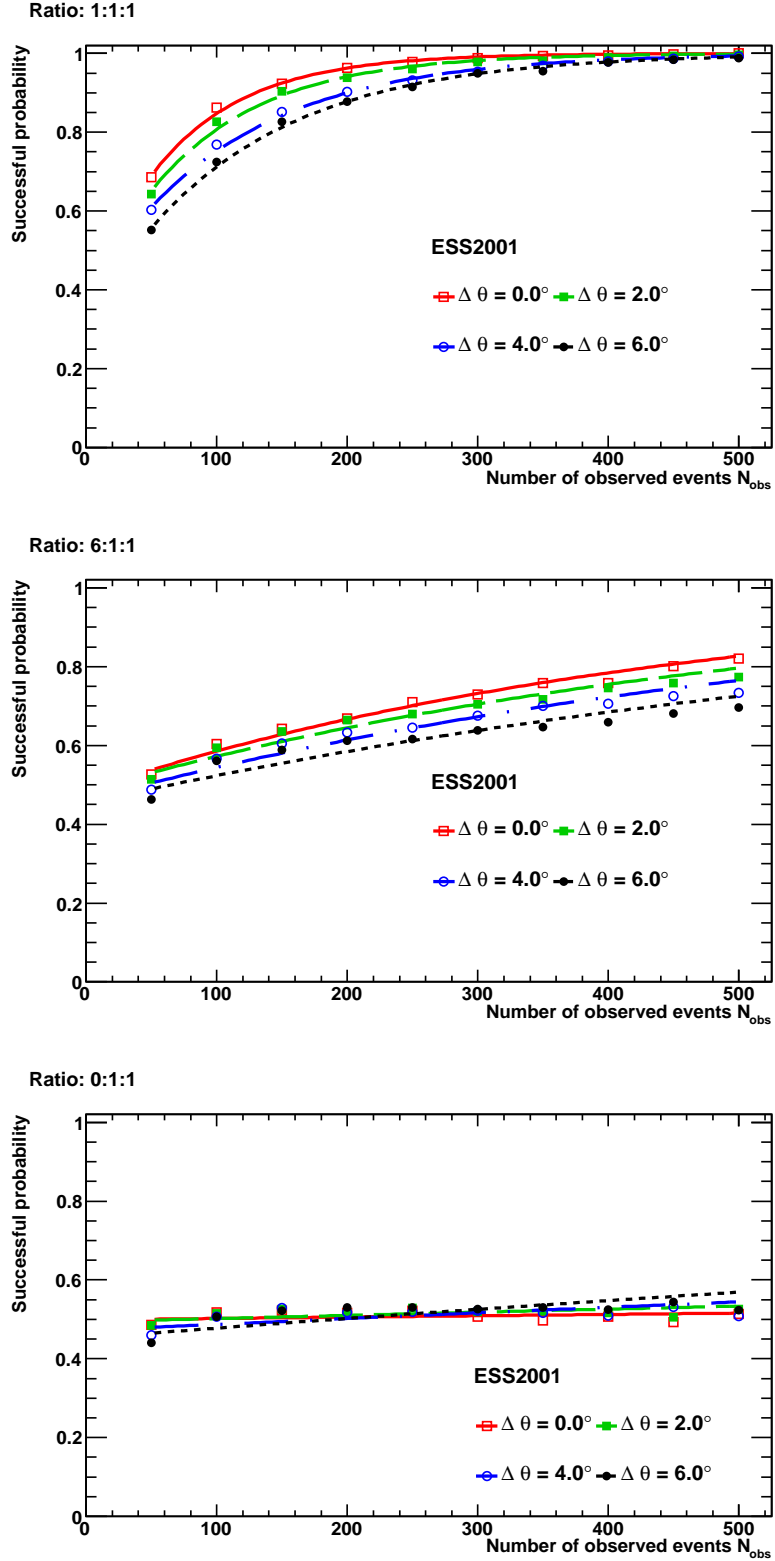}
        \caption{   \label{fig:ProbSuccess-ESS-2km}
            The successful probability of flavor ratio extraction as a function of the number of observed events $N_{\textrm{obs}}$, assuming initial ratio of $1/3:1/3:1/3$ (top panel), $0.75:0.125:0.125$ (middle panel), and $0:0.5:0.5$ (bottom panel), where $N_{\textrm{obs}}$ ranges from $50$ to $500$ with an interval of $50$. Results for different angular resolutions are plotted: $\Delta \theta=$ \SI{0}{\degree} (red), \SI{2}{\degree} (green), \SI{4}{\degree} (blue), \SI{6}{\degree} (black).
            %
            %
            %
            The ESS neutrino flux \cite{ESS2001} is assumed in both data generating and fitting, and the $\nu_{\mu}$-$\nu_{\tau}$ symmetry is assumed in fitting so that the relative ratio between them is known and fixed.
        }
     \end{figure*}

    After every data set in the ensemble has been fitted for a given $N_{\textrm{obs}}$ and $\Delta \theta$, the successful probability of flavor ratio extraction is calculated as the number of successful data sets divided by the total number of data sets.
    The result is shown in Fig.~\ref{fig:ProbSuccess-ESS-2km}.

    In general, the successful probability increases with the number of observed events and a better angular resolution, as expected.
    The probability for the standard scenario (top panel) is greater than those for the decay scenarios (the one with normal hierarchy is shown in the middle panel) since the latter have true $\nu_{e}$ ratios so closed to the boundaries that even small statistical fluctuation may result in failure.
    But the probability for initial ratio of $0:0.5:0.5$ is always around \SI{50}{\percent} regardless of the number of observed events and angular resolution assumed because its actual $\nu_{e}$ ratio is on the boundary.

    For the standard scenario, the successful probability is greater than \SI{70}{\percent} for $N_{\textrm{obs}} \geq 100$ and becomes over \SI{90}{\percent} for $N_{\textrm{obs}} \geq 250$, if the angular resolution is within \SI{6}{\degree}. On the other hand, for the decay scenario with normal hierarchy, the probability is over \SI{50}{\percent} once $N_{\textrm{obs}} \geq 100$.
    It was reported in Ref.~\cite{ARA2011} that fully deployed ARA is able to detect about 50 cosmogenic neutrinos in three years.
    %
    Therefore our method is feasible for ARA to extract the flavor ratio of cosmogenic neutrinos.

    We define the resolution of $\nu_{e}$ event fraction as the spread of all fitted values $\hat{f}_{e}$s in the ensemble with respect to the expected value $f_{e,\textrm{exp}}$,
    %
    \begin{equation}
        R_{\pm}(N_{\textrm{obs}}, \Delta\theta) =  \sqrt{ \frac{1}{ N_{\textrm{s}}-1 }  \sum_{i=1}^{N_{\textrm{s}}} (\hat{f}_{e,i}- f_{e,\textrm{exp}} )^{2} },
    \end{equation}
    where $N_{\textrm{s}}$ is the number of hypothetical experimental data sets.
    %
    %
    %
    If the fitting is not disturbed by the boundary cutoff, then the distribution of $\hat{f}_{e}$s asymptotically approaches a Gaussian distribution as the number of observed events $N_{\textrm{obs}}$ increases, and therefore the resolution defined here is approximately $1\sigma$ uncertainty at \SI{68}{\percent} confidence level.
    %
    In order to reduce the effect induced by boundary cutoff, the resolution is calculated separately at both sides of the expected value and denoted as $R_{+}$ and $R_{-}$.

%

\begin{figure}
         \includegraphics[width=0.7\linewidth]{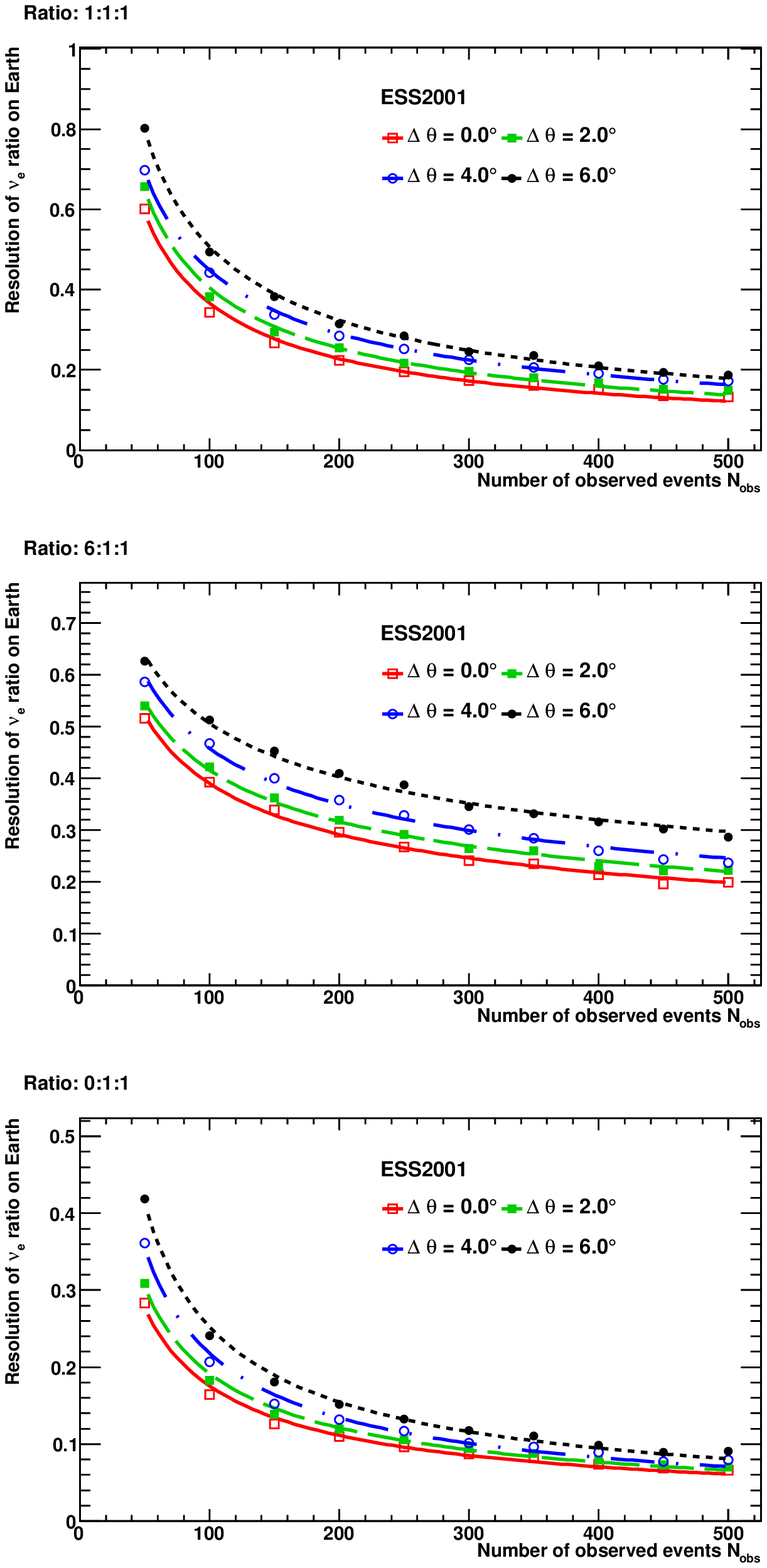}
        \caption{   \label{fig:FlavorResolution-ESS-2km-v2}
           The resolution of $\nu_{e}$ ratio at the Earth's surface $R^{E}$ (see text for definition) versus the number of observed events $N_{\textrm{obs}}$ for angular resolution of neutrino direction $\Delta\theta=$ \SI{0}{\degree} (the perfect case, red solid line and open square), \SI{2}{\degree} (green dashed line and full square), \SI{4}{\degree} (blue dash-dotted line and open circle), and \SI{6}{\degree} (black dotted line and full circle).
           Results assuming initial flavor ratios of $\frac{1}{3}:\frac{1}{3}:\frac{1}{3}$ (top panel), $0.75:0.125:0.125$ (middle panel), and $0:0.5:0.5$ (bottom panel) are shown.
           The ESS neutrino flux \cite{ESS2001} is assumed in both data generating and fitting, and the $\nu_{\mu}$-$\nu_{\tau}$ symmetry is assumed in fitting so that the relative ratio between them is known and fixed.
        }
 \end{figure}

    The upper and lower bounds of $\nu_{e}$ event fraction, $f_{e,\textrm{exp}} \pm R_{\pm}$, are then transformed into the corresponding flavor ratios at the Earth's surface, $f_{e,\textrm{true}}^{E} \pm R_{\pm}^{E}$, according to Eqn.~\ref{eq:RatioConversion} (with $f$ interchanging with $f^{E}$; see also Fig.~\ref{fig:eFlavorConversion-2km}).
    %
    %
    %
    The resolution of $\nu_{e}$ ratio on the Earth $R^{E}$ for given $N_{\textrm{obs}}$ and $\Delta \theta$ is plotted in Fig.~\ref{fig:FlavorResolution-ESS-2km-v2}. The lower (upper) resolution is taken in the decay scenario with normal (inverted) hierarchy. The resolution in the standard scenario is asymmetric on both sides, with the upper resolution worse than the lower one typically by about $0.06$, due to the nonlinear conversion relation between event ratios and flux ratios (Fig.~\ref{fig:eFlavorConversion-2km}), and hence the average value is taken.

\begin{figure*}
         \includegraphics[width=1\linewidth]{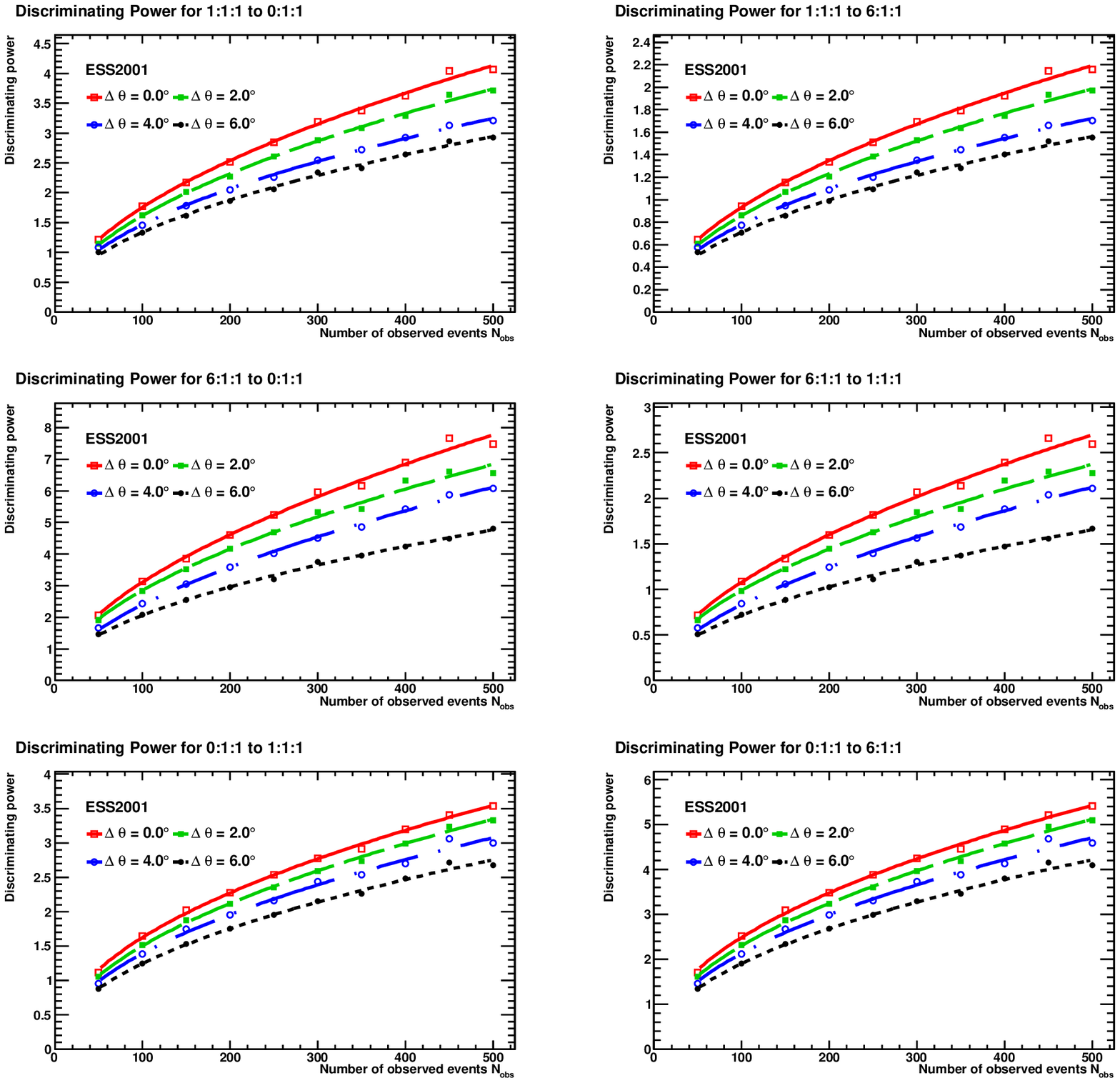}
        \caption{   \label{fig:DiscriminatingPower-ESS-2km}
           The discriminating power of $\nu_e$ flavor ratio versus the number of observed events $N_{\textrm{obs}}$ for angular resolution of neutrino direction $\Delta\theta=$ \SI{0}{\degree} (the perfect case, red solid line and open square), \SI{2}{\degree} (green dashed line and full square), \SI{4}{\degree} (blue dash-dotted line and open circle), and \SI{6}{\degree} (black dotted line and full circle).
           Results assuming initial flavor ratios of $1/3:1/3:1/3$ (top panels), $0.75:0.125:0.125$ (middle panels), and $0:0.5:0.5$ (bottom panels) are shown.
           The ESS neutrino flux \cite{ESS2001} is assumed in both data generating and fitting, and the $\nu_{\mu}$-$\nu_{\tau}$ symmetry is assumed in fitting so that the relative ratio between them is known and fixed.
        }
 \end{figure*}

    To measure the separability of a flavor ratio predicted in a ``fake'' scenario from that extracted from a given ``true'' scenario, we define the discriminating power of flavor ratios as
    \begin{equation}
      \textrm{discriminating power} \equiv \frac{|f_{e,\textrm{model} }-f_{e, \textrm{exp} }|}{R},
    \end{equation}
    %
    %
    where $f_{e,\textrm{model}}$ is the $\nu_{e}$ event ratio expected by the scenario to be examined, and $f_{e, \textrm{exp}}$ and $R$ are the expected $\nu_{e}$ event ratio and its resolution in the assumed true scenario, respectively.
    Note it is the $\nu_{e}$ ``event'' ratio and its resolution that are used in the definition because of the asymptotic normality of its distribution, whereas the distribution of $\nu_{e}$ flux ratio $f_{e}^{E}$ is skew due to the nonlinear conversion relation.
    The discriminating powers assuming different true scenarios are shown in Fig.~\ref{fig:DiscriminatingPower-ESS-2km}.

    In Ref.~\cite{ARA2011}, the ARA simulation result reports an angular resolution of neutrino direction about \SI{6}{\degree}. Under this circumstances, given $100$ observed neutrino events, the successful probability is about \SI{70}{\percent} in the standard scenario while about \SI{50}{\percent} for both decay scenarios (black curves in Fig.~\ref{fig:ProbSuccess-ESS-2km}).
    If the standard scenario is the real one, the discriminating power with respect to the decay scenario with normal (inverted) hierarchy will be at about $0.65\sigma$ ($1.3\sigma$) level.
    If the decay scenario with normal (inverted) hierarchy is real, then the discriminating power with respect to the standard scenario will be at about $0.7\sigma$ ($1.2\sigma$) level, while that to its inverted (normal) counterpart will be at about $2.0\sigma$ ($1.8\sigma$) level.
    Hence a preliminary constraint on the neutrino decay can be set by the method proposed in this article.



%

    %
    If the number of the observed events is doubled with the angular resolution fixed ($N_{\textrm{obs}}=200$, $\Delta \theta=$ \SI{6}{\degree}), the successful probability rises to \SI{85}{\percent} for the standard scenario while still around \SI{50}{\percent} for the decay ones; the discriminating power between the standard scenario and either of the decay scenario increases by about $0.3$--$0.5\sigma$ while by about $0.8$--$1\sigma$ between the decay scenarios.
    If the angular resolution is improved by \SI{4}{\degree} with the number of observed events fixed ($N_{\textrm{obs}}=100$, $\Delta \theta=$ \SI{2}{\degree}), the successful probability becomes about \SI{80}{\percent} for the standard scenario while still around \SI{50}{\percent} for both decay ones; and the discriminating power between the standard scenario and either of decay one increases by about $0.2$--$0.3\sigma$, and by about $0.4\sigma$ between the decay scenarios.
    The limit of our method with perfect angular resolution is also shown in Fig.~\ref{fig:ProbSuccess-ESS-2km}, Fig.~\ref{fig:FlavorResolution-ESS-2km-v2} and Fig.~\ref{fig:DiscriminatingPower-ESS-2km} (red curves).

%


    To accumulate more neutrino events, apart from waiting for more neutrinos to come, one can increase the event rate by extending the antenna array. To achieve a higher angular resolution of neutrino direction, on the other hand, an antenna array with denser grid is required for a better imaging of the Cherenkov cone. Therefore the results presented here can serve as a reference for optimizing the future detector configuration of ARA or other similar observatories.



    \subsection{ Systematic Uncertainty from Neutrino Spectrum }

\begin{figure}
         \includegraphics[width=0.7\linewidth]{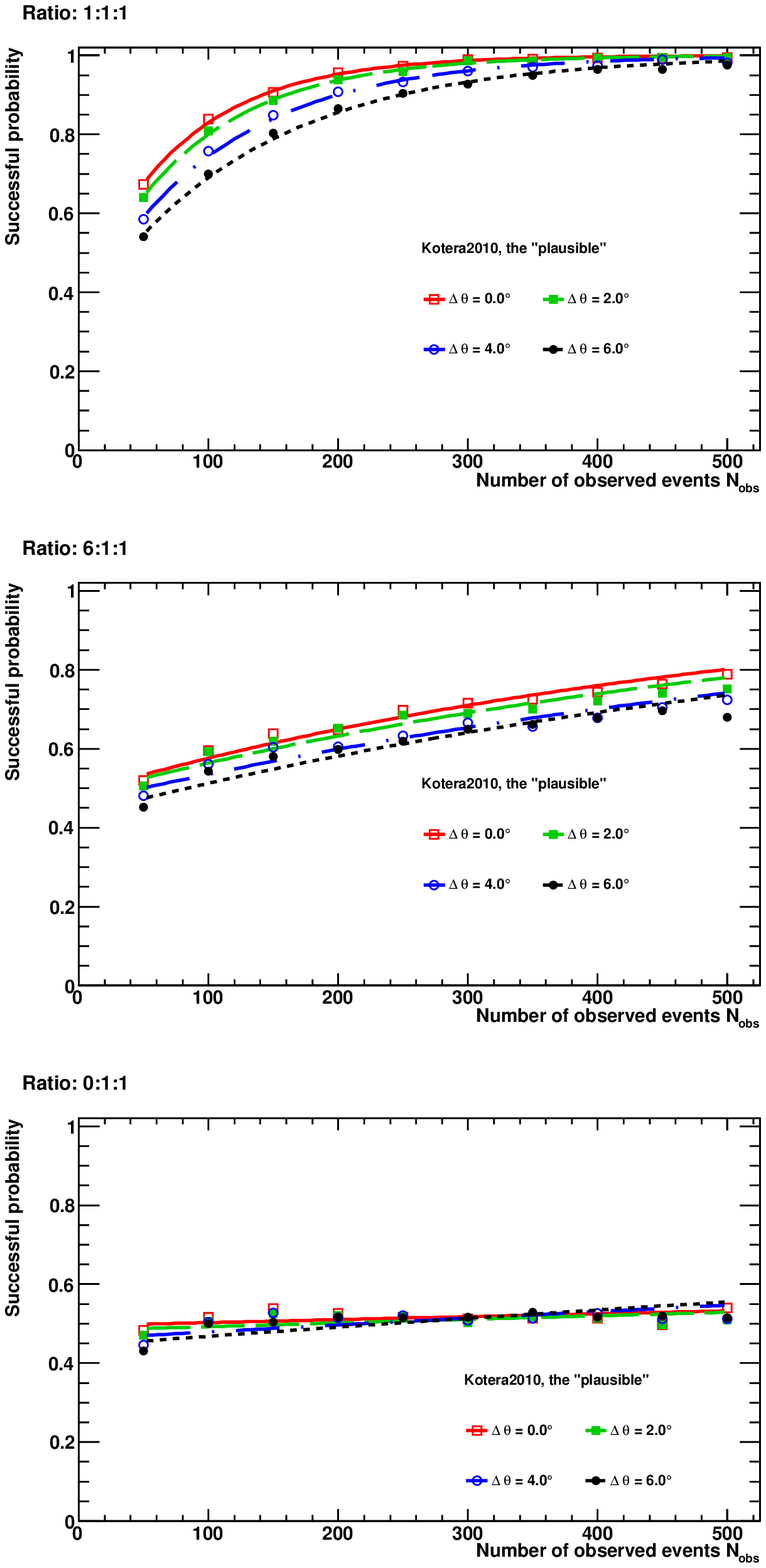}
        \caption{   \label{fig:ProbSuccess-Kotera-2km}
            The successful probability of flavor ratio extraction. Similar to Fig.~\ref{fig:ProbSuccess-ESS-2km} except that the assumed neutrino flux is the plausible scenario in Ref.~\cite{Kotera2010} (blue dashed curve in Fig.~\ref{fig:GZKFlux}).
        }
     \end{figure}

\begin{figure}
         \includegraphics[width=0.7\linewidth]{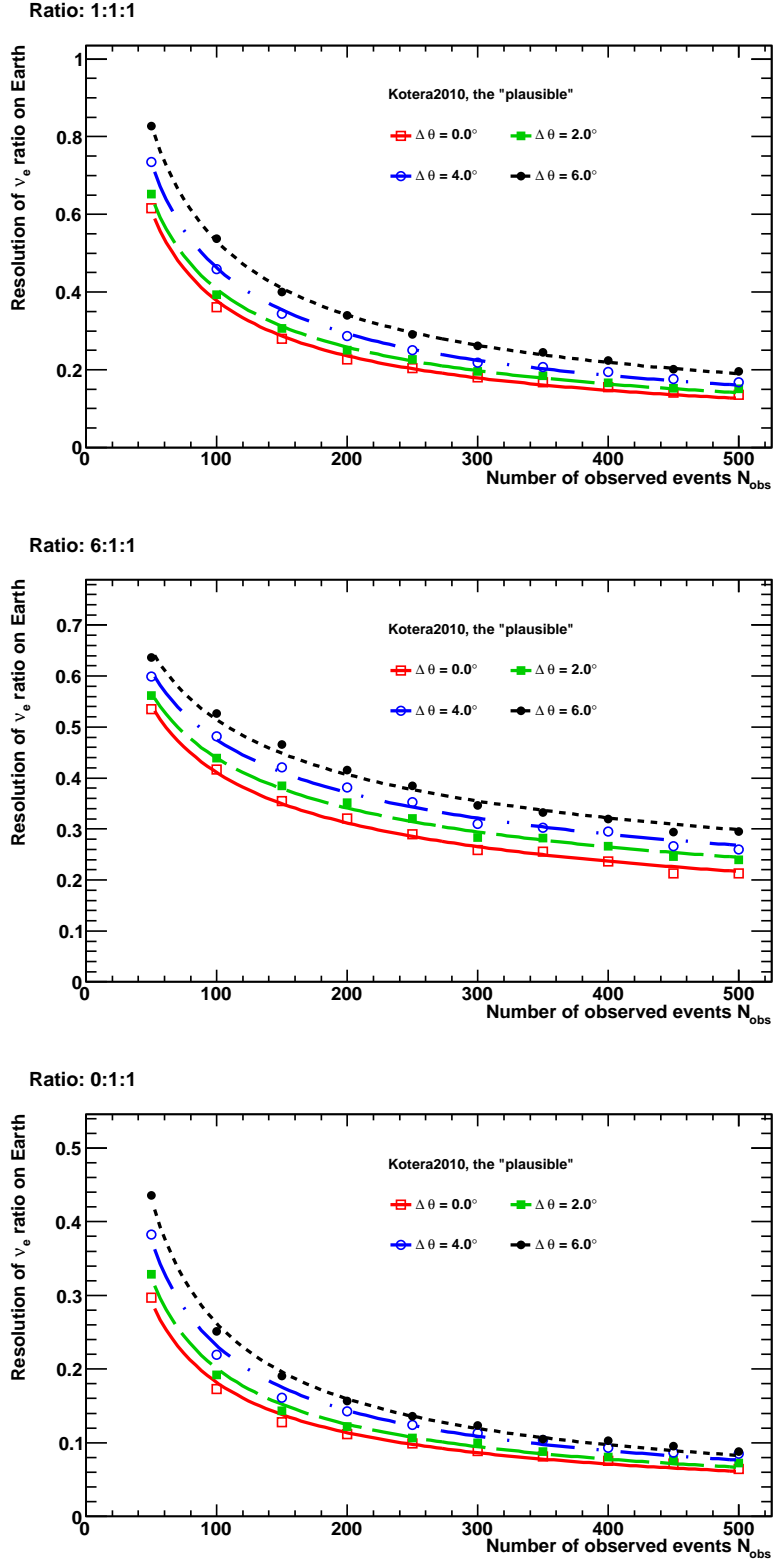}
        \caption{   \label{fig:FlavorResolution-Kotera-2km-v2}
           The resolution of $\nu_{e}$ ratio at the Earth's surface versus the number of observed events. Similar to Fig.~\ref{fig:FlavorResolution-ESS-2km-v2} except that the assumed neutrino flux is the plausible scenario in Ref.~\cite{Kotera2010} (blue dashed curve in Fig.~\ref{fig:GZKFlux}).
        }
 \end{figure}

\begin{figure}
         \includegraphics[width=1\linewidth]{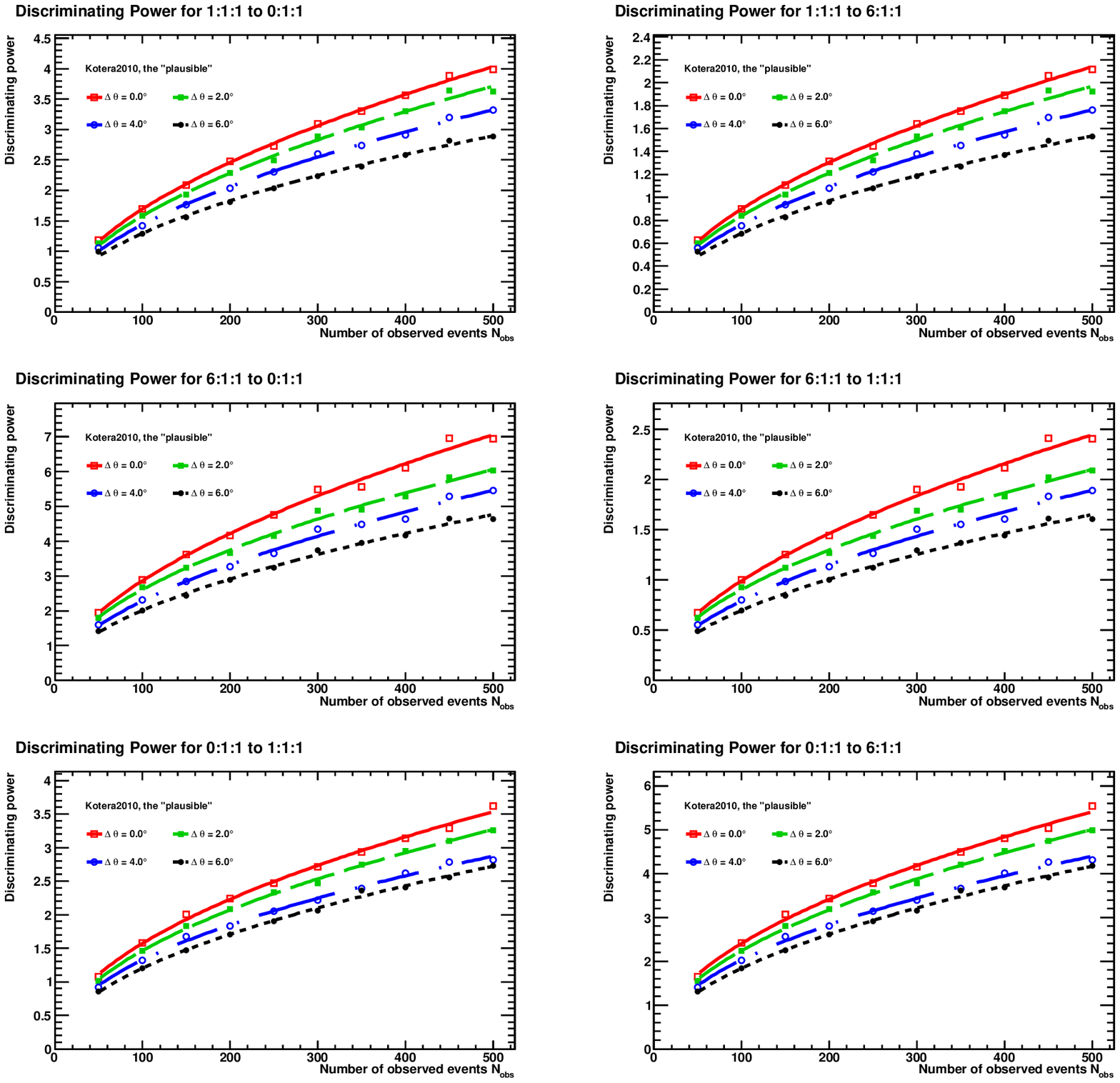}
        \caption{   \label{fig:DiscriminatingPower-Kotera-2km}
           The discriminating power of $\nu_{e}$ flavor ratio versus the number of observed events. Similar to Fig.~\ref{fig:DiscriminatingPower-ESS-2km} except that the assumed neutrino flux is the plausible scenario in Ref.~\cite{Kotera2010} (blue dashed curve in Fig.~\ref{fig:GZKFlux}).
        }
 \end{figure}

    Fig.~\ref{fig:ProbSuccess-Kotera-2km}, Fig.~\ref{fig:FlavorResolution-Kotera-2km-v2} and Fig.~\ref{fig:DiscriminatingPower-Kotera-2km}
    show the results assuming the neutrino flux predicted in the ``plausible'' scenario in Ref.~\cite{Kotera2010} (blue dashed curve in Fig.~\ref{fig:GZKFlux}).
    %
    %
    The successful probability is lower than that assuming the ESS flux and the resolution is worse. Recall that the flux in Ref.~\cite{Kotera2010} is steeper than the ESS one, i.e.~there are more neutrino events with lower energies. Since the interaction probabilities between flavors are less distinct from each other at lower energies (see Fig.~\ref{fig:NuIntProb}), the event direction distributions between flavors are harder to be distinguished.


    If the observer does not exactly know the ``real'' neutrino spectrum (i.e., the one used to generate pseudo-data), then the mismatch in shape between the real and the expected (i.e., the one used to fit the data) spectra will introduce a systematic bias to the extracted ratio. We found that if the real spectrum is the plausible scenario in \cite{Kotera2010} while the spectrum expected by the observer is the ESS model, the bias is about \SI{10}{\percent}. More specifically, the average fitted value drops from 0.58 to 0.53, because a steeper neutrino spectrum leads to more events coming from below the horizon due to the longer interaction length of neutrinos with lower energies.
    In order to fit the event distribution from the steeper spectrum with the flatter one, the $\nu_{\tau}$ fraction has to be increased while the $\nu_{e}$ fraction decreased.


    \subsection{Systematic Uncertainty from Neutrino Cross Sections}

\begin{figure}
         \includegraphics[width=0.7\linewidth]{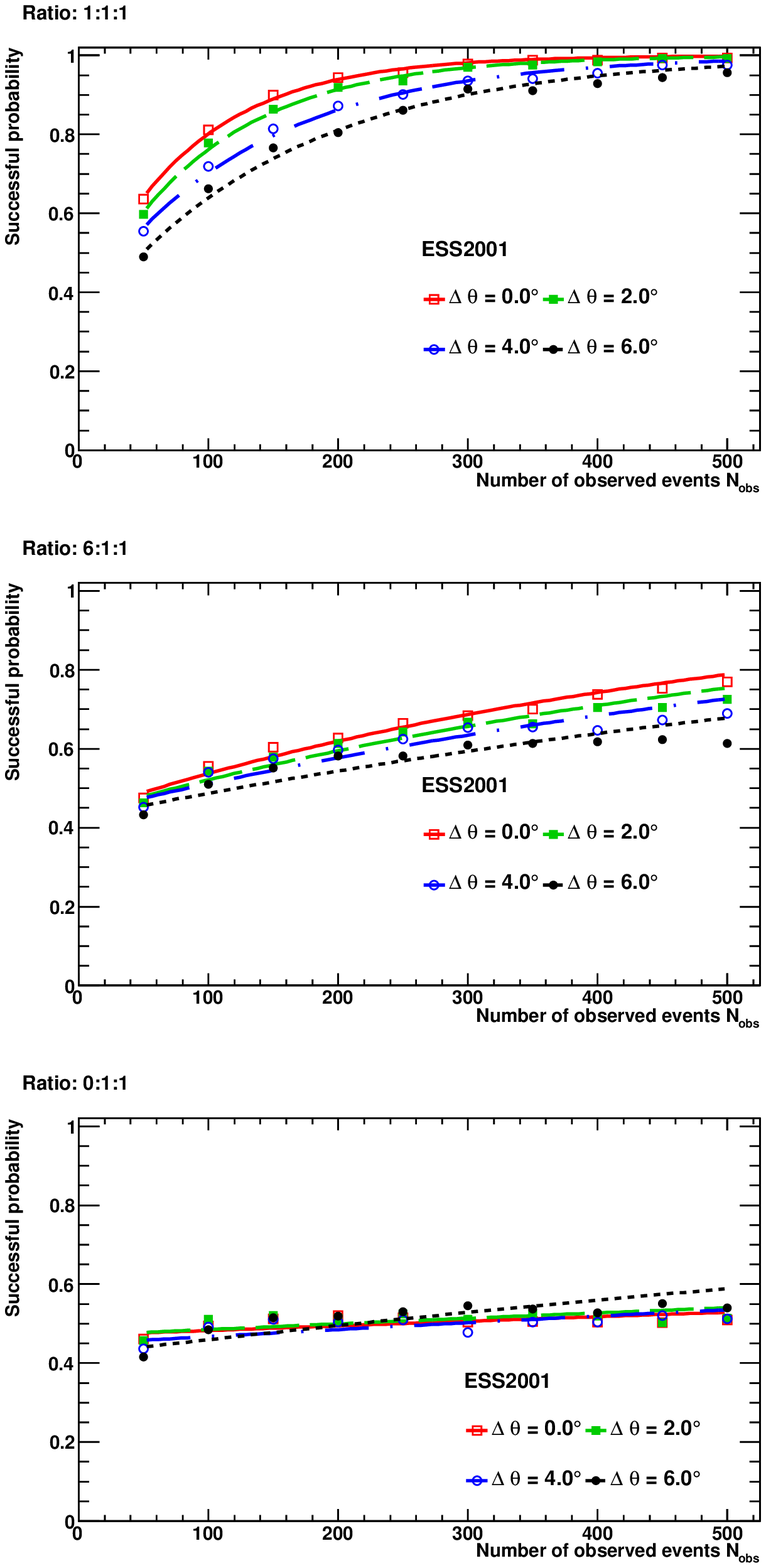}
        \caption{   \label{fig:ProbSuccess-ESS-2km-x2}
            The probability of successful flavor ratio reconstruction. Same as Fig.~\ref{fig:ProbSuccess-ESS-2km} except that the neutrino cross section is multiplied by a factor of two.
        }
     \end{figure}

\begin{figure}
         \includegraphics[width=0.7\linewidth]{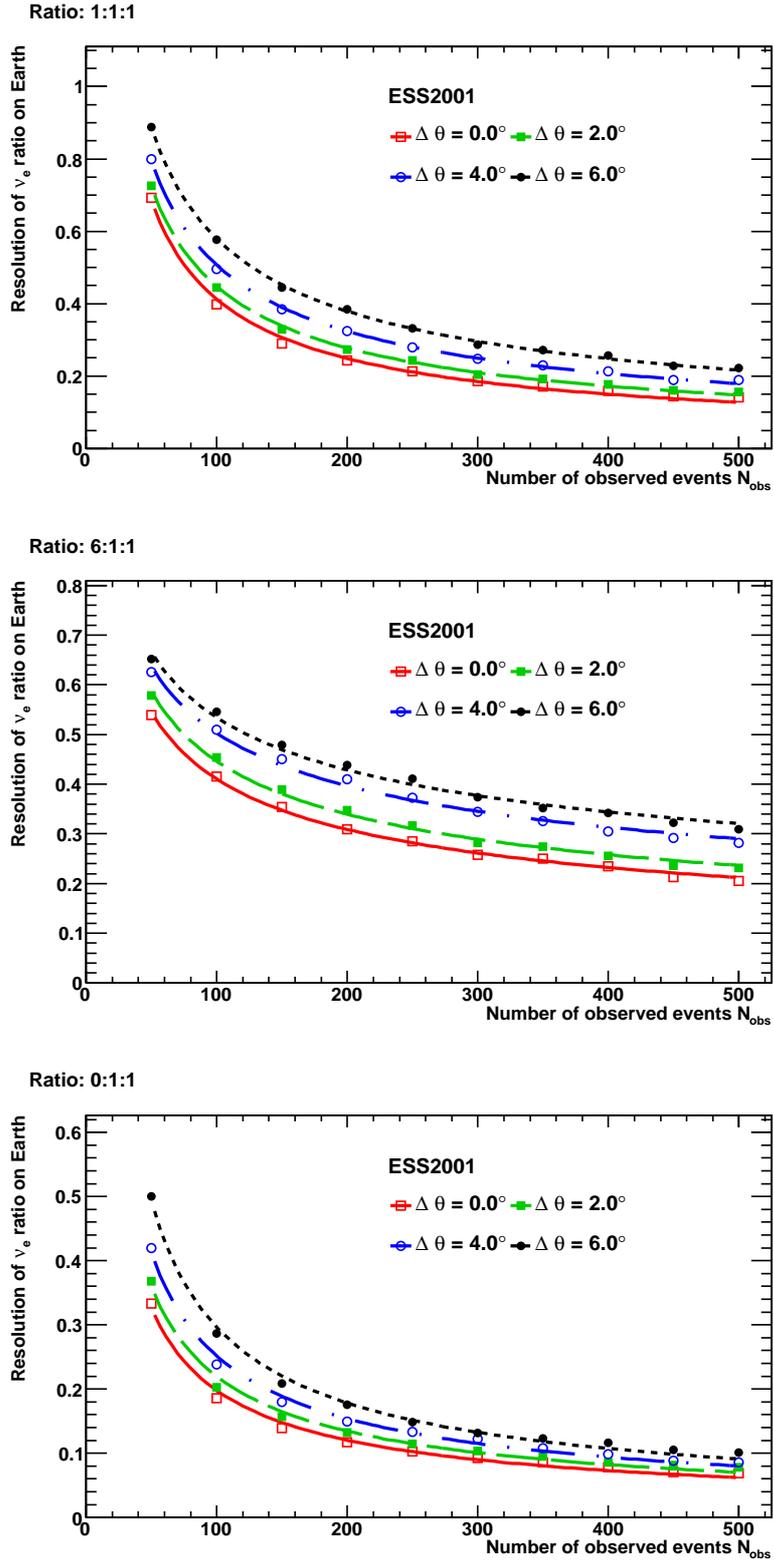}
        \caption{   \label{fig:FlavorResolution-ESS-2km-v2-x2}
           The resolution of $\nu_{e}$ ratio on the Earth. Same as Fig.~\ref{fig:FlavorResolution-ESS-2km-v2} except that the neutrino cross section is multiplied by a factor of two.
        }
 \end{figure}

\begin{figure*}
         \includegraphics[width=1\linewidth]{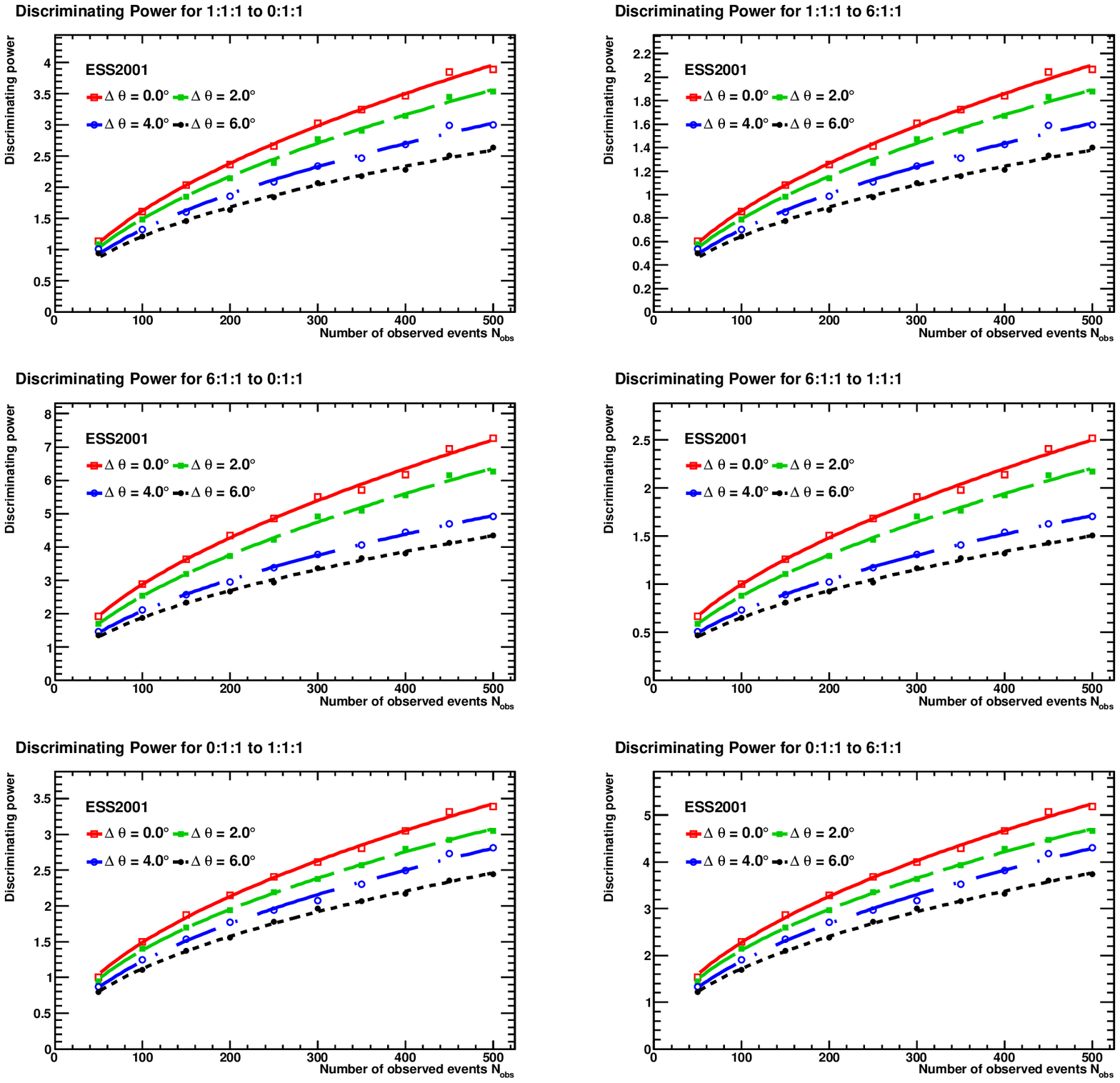}
        \caption{   \label{fig:DiscriminatingPower-ESS-2km-x2}
           The discriminating power of $\nu_{e}$ flavor ratio. Same as Fig.~\ref{fig:DiscriminatingPower-ESS-2km} except that the neutrino cross section is multiplied by a factor of two.
        }
 \end{figure*}

    The calculation of neutrino cross section in Ref.~\cite{GQRS1998} reports an uncertainty of factor of $2^{\pm1}$ at around \SI{e20}{\eV}. Therefore we vary the neutrino cross section by a factor of two, and the results are shown in Fig.~\ref{fig:ProbSuccess-ESS-2km-x2}, Fig.~\ref{fig:FlavorResolution-ESS-2km-v2-x2} and Fig.~\ref{fig:DiscriminatingPower-ESS-2km-x2}.
    %
    %
     A larger neutrino cross section leads to lower successful probability and resolution of flavor ratio, because the shorter interaction length increase the fraction of events in the down-going directions where the interaction probabilities for different flavors are more or less the same, while makes the more distinct, up-going parts slightly decrease.

\section{Conclusion and Future Works}

    In summary, measuring the flavor ratio of UHE cosmic neutrinos can not only reveal the physical conditions at UHECR sources but also probe neutrino oscillation parameters and non-standard physics that might involve during the propagation. Neutrino observatories using the radio Cherenkov technique such as ARA \cite{ARA2011}, ARIANNA \cite{ARIANNA2007, *ARIANNA2010}, and SalSA \cite{SalSA2002}, are sensitive in the UHE regime and expected to accumulate of order of 10 to 100 cosmogenic neutrinos per year in the near future, and hence provide sufficient statistics for the flavor ratio identification.


    In this work, the direction distribution of neutrino events is proposed to determine the flavor ratio. In order to investigate its feasibility, a simulation is constructed and the expected distribution for ARA is derived. It is found that $\nu_{\tau}$ distribution has a different shape from the other two, but the distributions for $\nu_{e}$ and $\nu_{\mu}$ resemble each other. Therefore an additional constraint, the $\nu_{\mu}$-$\nu_{\tau}$ symmetry, is imposed on the fitting of data distribution to avoid multiple solutions.


    This method is proved to be feasible for ARA, e.g., for 100 events and an angular resolution of \SI{6}{\degree}, the successful probability is about \SI{70}{\percent} in the standard scenario and over \SI{50}{\percent} for the neutrino decay models considered here.
    This method is also able to set a preliminary constraint on the neutrino decay, e.g., given an angular resolution of \SI{6}{\degree}, it requires about 250 events to separate the standard scenario from the decay model with inverted hierarchy by about $2\sigma$ level.
    %
    Therefore, the flavor resolution as a function of the number of observed events and the angular resolution of neutrino direction presented here can serve as a reference for optimizing the future configuration of ARA. Similar procedures can be done for other observatories.



    However, we have not fully taken advantage of all the information possessed by neutrino events. For example, the spatial distribution and the energies of showers, which can be retrieved by incorporating the vertex reconstruction, can help to distinguish electron flavor from the other two; and the characteristics of radio signals may allow the classification of shower types and further the identification of the neutrino flavor event by event.
    In addition, the near-field effect of Cherenkov radiation has to be considered, since in some cases the observation distances are comparable to the shower lengths and the far-field approximation applied here would underestimates both the signal strength and the angular width of the Cherenkov cone, as pointed out in Ref.~\cite{Hu2012}.
    %
    The neutrino spectrum and the cross sections should also be incorporated into the fitting in order to reduce the systematic uncertainties.
    These will be included in the future simulation, and we expect to get a higher resolution of the neutrino flavor ratio.





\begin{acknowledgments}
We thank David Seckel and his coworkers of University of Delaware for sharing with us their simulation code SADE; our investigation would have been incomplete without this code. We also thank Guey-Lin Lin of National Chiao-Tung University for valuable discussions and suggestions. This work is supported by Taiwan National Science Council under Project No.~NSC100-2119-M-002-025- and by US Department of Energy under Contract No.~DE-AC03-76SF00515.
\end{acknowledgments}


%

\end{document}